\documentclass[sigconf, nonacm]{acmart}
\usepackage{inputenc}
\usepackage{newtxmath}
\usepackage{mathtools}
\usepackage[noend]{algpseudocode}
\usepackage{algorithm,algorithmicx}
\usepackage{graphicx}
\usepackage{booktabs}
\usepackage{array}
\usepackage{adjustbox}
\usepackage[belowskip=-15pt,aboveskip=2pt]{caption}
\usepackage{listings}
\usepackage{url}
\usepackage{color}
\usepackage{xcolor}
\usepackage{enumerate}
\usepackage{lipsum}
\usepackage{pbox}
\usepackage{wasysym}
\usepackage{subfig}
\usepackage{float}
\usepackage{paralist}
\usepackage[inline, shortlabels]{enumitem}
\usepackage{subfiles}
\usepackage[title]{appendix}
\usepackage{atbegshi,afterpage}
\usepackage{balance}
\let\thepage\relax

\newcommand\vldbpagestyle{plain}

\newenvironment{ttmath}
 {\everymath{\ttmathgroup}\everydisplay{\ttmathgroup}}
 {}

\AtBeginDocument{%
  $\mathtt{\xdef\ttmathgroup{\fam\the\fam\relax}}$%
    \AtBeginShipoutNext{\AtBeginShipoutDiscard}
}

\definecolor{darkgreen}{rgb}{0.0,0.50,0.00}
\definecolor{darkblue}{rgb}{0.0, 0.0, 0.50}

\newcommand{\longpaper}[1]{}

\newcommand{\old}[1]{}
\newcommand{\movedfrom}[1]{}

\newcommand{\shorten}[1]{}

\newcommand{\shortRemove}[1]{}
\newcommand{\vsRemove}[1]{}

\newtheorem{example}{Example}
\newtheorem{definition}{Definition}
\newtheorem{theorem}{Theorem}
\newtheorem{proposition}{Proposition}

\begin{document}
\title{iWarded: A System for Benchmarking Datalog+/- Reasoning (technical report)}

\author{Teodoro Baldazzi}
\affiliation{%
  \institution{Universit\`a Roma Tre}
}
\author{Luigi Bellomarini}
\affiliation{%
  \institution{Banca d'Italia}
}
\author{Emanuel Sallinger}
\affiliation{%
  \institution{University of Oxford\\ and TU Wien}
}
\author{Paolo Atzeni}
\affiliation{%
  \institution{Universit\`a Roma Tre}
}

\begin{abstract}
Recent years have seen increasing popularity of logic-based reasoning systems, with research and industrial interest as well as many flourishing applications in the area of Knowledge Graphs. Despite that, one can observe a substantial lack of specific tools able to generate nontrivial reasoning settings and benchmark scenarios. As a consequence, evaluating, analysing and comparing reasoning systems is a complex task, especially when they embody sophisticated optimizations and execution techniques that leverage the theoretical underpinnings of the adopted logic fragment. In this paper, we aim at filling this gap by introducing iWarded, a system that can generate very large, complex, realistic reasoning settings to be used for the benchmarking of logic-based reasoning systems adopting Datalog+/-, a family of extensions of Datalog that has seen a resurgence in the last few years. In particular, iWarded generates reasoning settings for Warded Datalog+/-, a language with a very good tradeoff between computational complexity and expressive power. In the paper, we present the iWarded system and a set of novel theoretical results adopted to generate effective scenarios. As Datalog-based languages are of general interest and see increasing adoption, we believe that iWarded is a step forward in the empirical evaluation of current and future systems. 
\end{abstract}

\maketitle
\pagestyle{\vldbpagestyle}

\section{Introduction}
The role of knowledge as a driving economic force and the noteworthy growth in the amount of data stored and used as decision-making source over the last decade caused an increasing interest of companies in data, including how to manage and reason over it.

In parallel to the development of new methodologies for data processing, numerous benchmarking solutions have been created with the purpose of evaluating the performance of various systems, their limits and features, in order to guarantee appropriate levels of efficiency and scalability.Benchmarking is, in general, the process of running a specific program or workload on a specific machine or system and measuring the resulting performance~\cite{saavedra1996analysis}. As Patterson~\cite{patterson2012technical} states, ``When a field has good benchmarks, we settle debates and the field makes rapid progress''.  

The database community identifies distinct categories of benchmarking systems: among them, the canonical TPC benchmarks~\cite{tpcwebpage}, the standard option to evaluate database systems, and advanced research tools like {\sc iBench}, a schema mapping generator for the analysis of data integration and data exchange scenarios~\cite{arocena2015ibench}.

Yet, for Knowledge Graph Management Systems (KGMSs), and in particular Datalog$^\pm$-based KGMSs, hardly any benchmark exists. This makes precisely those gains and rapid advances observed in other parts of the data processing spectrum hard or impossible to obtain. Despite the importance of this field, only recently did Benedikt et al.~\cite{benedikt2017benchmarking} propose a set of query answering benchmarks with specific focus on {\sc chase}-based techniques~\cite{FKMP03}.

\smallskip
\noindent 
\textbf{Benchmarking Datalog$^\pm$ Reasoning}. 
This paper answers the need of providing a system able to generate  reasoning settings tailored to evaluate state-of-the-art reasoners using languages from the Datalog$^\pm$ family~\cite{CaGL09}, a broadly adopted logic-based reasoning framework experiencing relevant success for KGs. Our particular focus is the eponymous \emph{Warded} Datalog$^\pm$, which is able to express full Datalog as well as SPARQL under the OWL 2 QL entailment regime and set semantics. At the same time, it is \texttt{PTIME} complete in data complexity, thus providing a good balance of expressivity and computational complexity~\cite{gottlob2015beyond}.

In particular, this paper presents the {\sc iWarded} system, a benchmarking system for Warded Datalog$^\pm$. On the one hand, {\sc iWarded} meets the need of being a benchmark generator for Warded Data\-log$^\pm$: it allows to evaluate and compare reasoning strategies, optimization techniques, subtle tuning of systems that implement Warded Datalog$^\pm$, in particular the Vadalog system~\cite{bellomarini2018vadalog}. On the other hand and more generally, it provides tools for allowing other Datalog$^\pm$ systems to add Warded Datalog$^\pm$ to their repertoire, and offers a way to compare the performance of these systems. 

\smallskip
\noindent
\textbf{Benchmark Approach}. Adapting to our setting an observation made by Gray~\cite{gray1992benchmark}, 
a ``good'' benchmark should follow four criteria: (1)~\textit{relevance}: the benchmark should deal with real performance scenarios of the language; (2)~\textit{reusability}: the generated programs should be reusable and adaptable to test different aspects of the language; (3)~\textit{simplicity}: the benchmark should be feasible and should not require too many resources; (4)~\textit{scalability}: the benchmark should adapt to requests for scenarios of different dimensions and complexity. Along these lines, a benchmarking system for Warded Datalog$^\pm$ should allow to efficiently generate very large, complex, yet realistic scenarios with different characteristics. Each generated program should include multiple relevant aspects and properties of the language to build a foundation for a more elaborate and long-lasting process of analysis.

\smallskip
\noindent
\textbf{Vadalog Normalizer}. 
Because of the interaction of recursion and existential quantification in the languages (technically, ``fragments'') of the Datalog$^\pm$ family, reasoning systems implement specialized algorithms to exploit the full potential of the fragments and ensure termination in practice. This is the case of Vadalog, which exploits the restrictions of Warded Datalog$^\pm$~\cite{bellomarini2018swift}. As described in detail in~\cite{bellomarini2020vadalog,bellomarini2018vadalog}, the key theoretical underpinnings of \textit{Warded Datalog}$^\pm$ consist in a  form of \textit{reasoning boundedness}, which allows for efficient recursion control techniques that guarantee termination in practice while ensuring small memory footprint. Analogous boundedness results exist for the other Datalog$^\pm$ fragments~\cite{CaGL09,berger2019space}.

To be exploited, such reasoning boundedness requires the reasoning setting at hand to be in a ``\textit{normalized form}'': when it does not contain a specific kind of ``harmful'' joins between variables affected by existential quantification. In this case, we say the setting is expressed in \textit{Harmless Warded Datalog$^\pm$}. As a consequence, reasoners wanting to support the full extent of Warded Datalog$^\pm$ are required to implement such normalization capabilities.

To aim at the generation of unbiased benchmarks decoupling reasoning challenges from normalization, {\sc iWarded} includes a novel normalization algorithm (named the \textit{Vadalog normalizer} and incorporating \textit{Harmful Join Elimination}), based on original theoretical results we present for Warded Datalog$^\pm$.

\smallskip
\noindent
\textbf{Main Contributions}.
The main contribution of this paper is the  \textbf{\textsc{iWarded}} benchmark system, consisting of:
\begin{itemize}[leftmargin=2mm]\itemsep-\parsep
    \item the \textbf{\textsc{iWarded}} \textbf{benchmark generator}, a new tool to generate Vadalog benchmarks stimulating a wide range of language characteristics and features, its operating principles and the algorithms at its basis in the reasoner. We discuss the goals and the importance of the generator.
    \item the  \textbf{Vadalog normalizer}, a tool including a new theoretical basis for Warded Datalog$^\pm$ normalization as well as the full \textit{Harmful Join Elimination} (HJE) algorithm.
    \item an \textbf{experimental evaluation} showing how {\sc iWarded} is operated in practice and highlighting the characteristics of the benchmarks. The results are complemented with comparisons with the {\sc iBench} system, a benchmark generator that while not designed to generate Warded Datalog$^\pm$, is the closest ``sibling'' to {\sc iWarded}.
\end{itemize}

\smallskip
\noindent
\textbf{Organization}. The rest of the paper is structured as follows. Section~\ref{vadalog} describes the main characteristics of Vadalog, with particular focus on all the elements of the language that are used in this work. Section~\ref{iWarded} illustrates {\sc iWarded}, from the algorithm at its basis to its main use case, and discusses the process of benchmark generation. Section~\ref{hje} describes the theoretical problems behind Harmful Join Elimination, presents our theoretical results and algorithm. Section~\ref{testing} provides the experimental results with {\sc iWarded} for the generation of Vadalog programs and the comparison with {\sc iBench} as well as relevant experimental evaluations regarding the elimination of harmful joins. Section~\ref{relatedWork} discusses related work and in Section~\ref{conclusion} we draw up our conclusions.
\label{introduction}
\section{The Vadalog Language}
\noindent
To guide our discussion, in this section we briefly illustrate the main aspects and characteristics of \textit{Warded Datalog}$^\pm$, the logical core of the Vadalog language. A detailed analysis of the language as well as the theoretical implications can be found in~\cite{gottlob2019vadalog}.

\noindent
\textbf{Existentials and Reasoning}.
Datalog$^\pm$ extends Datalog~\cite{ceri1989you} with existential quantification in the rule conclusion and other features (the + in {$\pm$}) to make it suitable for ontological reasoning, while restricting the syntax in order to ensure scalability, decidability and data tractability (the - in {$\pm$}). A \textit{rule} is a first-order sentence of the form
\(
\forall \bar x \forall \bar y (\varphi(\bar x,\bar y)\ \rightarrow\ \exists \bar z \, \psi(\bar x, \bar z))
\), where $\varphi$ (the {\em body}) and $\psi$ (the {\em head}) are conjunctions of atoms.
As usual in this context, we omit universal quantifiers and denote conjunction by comma. An alternate syntax for rules is \({\ttmath
 \psi(\bar x, \bar z)~{\textnormal{:-}}~\varphi(\bar x,\bar y)
}\), adopting right-to-left implications, ``:-'' for ``$\leftarrow$'', and omitting existential quantifiers.

The semantics of a set of existential rules $\Sigma$ over a database $D$ is defined via the {\sc chase} procedure~\cite{MaMS79}, denoted as $\Sigma$(D), shown in Example~\ref{ex:mother}: the chase adds new atoms to $D$, possibly involving freshly generated symbols, namely \textit{labelled} or \textit{marked} \textit{nulls} for satisfying the existentially quantified variables, until $\Sigma(D)$ satisfies all the existential rules.

\begin{example}
\label{ex:mother} Consider the database $D = \{{\rm Person}({\rm Alice})\}$, and the set of existential rules
\begin{center}
$\textnormal{Person}(x)\to\exists{z}\,\textnormal{HasMother}(x,z).\textnormal{HasMother}(x,y)\to\textnormal{Person}(y).$
\end{center}
The database atom triggers the above existential rule, the chase starts and adds the following facts to $D$, where $\nu_1$ is a labelled null. 
\begin{center}
    $\textnormal{HasMother}({\rm Alice},\nu_1)\quad {\rm and}\quad \textnormal{Person}(\nu_1)$
\end{center}
The new fact \textnormal{Person}$(\nu_1)$ triggers again the existential rule, and the chase adds the facts
\begin{center}
    $\textnormal{HasMother}(\nu_1,\nu_2)\quad {\rm and} \quad \textnormal{Person}(\nu_2)$
\end{center}
where $\nu_2$ is a new labelled null. The result of the chase is the instance
\begin{gather*}
\{\textnormal{Person}({\rm Alice}),\textnormal{HasMother}({\rm Alice},\nu_1)\}\enskip\cup \\
\cup_{i>0}\{\textnormal{Person}(\nu_i),\textnormal{HasMother}(\nu_i,\nu_{i+1})\},
\end{gather*}
where $\nu_1$,$\nu_2$,... are labelled nulls.
\end{example}

\noindent
Given a database $D$ and a pair $Q=(\Sigma,Ans)$, where $\Sigma$ is a set of rules and $Ans$ an \textit{n}-ary predicate, we define the evaluation of $Q$ over $D$ as the set of tuples $Q(D,\Sigma)=\{\bar{t}\in dom(D)^n\,|\,Ans(\bar{t})\in \Sigma(D)\}$, where $\bar{t}$ is a tuple of constants. We denote \textit{universal tuple inference} (or simply \textit{reasoning task}) as the task of finding a database instance $J$ such that: (i)~$\bar{t}\in J$ if and only if Ans($\bar{t})\in Q(D,\Sigma)$ and (ii)~for every other instance $J^\prime$ such that $\bar{t}\in J^\prime$ if and only if $\bar{t}\in Q(D,\Sigma)$, there is a homomorphism $h$ from $J$ to $J^\prime$~\cite{bellomarini2020vadalog}.

\smallskip
\noindent
\textbf{Wardedness}. Warded Datalog$^\pm$ introduces syntactic conditions that constrain the propagation of labelled nulls in the reasoning task, by isolating the \textit{frontier variables} (i.e., universally quantified and appearing in the head) that could possibly bind to labelled nulls to appear in one specific body atom (the ward). 

Given a set of rules $\Sigma$, a position $\pi[i]$ (i.e., the $i$-th term of an atom $\pi$, with $=1,\ldots$) is \textit{affected} if (i)~$\pi$ appears in a rule $\rho$ of $\Sigma$ and $\pi[i]$ contains an existentially quantified variable (${\rm HasMother}[2]$ in Example~\ref{ex:mother} is affected) or, (ii)~ there is a rule $\rho$ of $\Sigma$ s.t.\ a frontier variable $x$ of $\rho$ only appears in affected body positions and in position $\pi[i]$ in the head (${\rm Person}[1]$ is affected).

A variable $x$ is \textit{harmless} with respect to a rule, if $x$ appears in a non-affected position, otherwise it is \textit{harmful}; a rule that contains a harmful variable is a \textit{harmful rule}, otherwise it is \textit{harmless rule}. A frontier harmful variable $x$ is called \textit{dangerous}. Intuitively, dangerous variables allow labelled nulls to propagate. For example, the variable \textit{y} in the second rule of Example~\ref{ex:mother} is dangerous. Rules containing dangerous variables are named \textit{dangerous rules}.

A rule is \textit{warded} if the following conditions hold: (i)~all the dangerous variables appear only in a single body atom, the \textit{ward}; and (ii)~a variable $x$ of the ward appears in another body atom iff it is harmless. A set of rules $\Sigma$ is warded if all its rules are warded. In Example~\ref{ex:mother}, the set of rules is warded, as ${\rm Person}$ in the body of the first and ${\rm HasMother}$ in the body of the second rule are wards.
\label{vadalog}
\section{{\lowercase{i}Warded}: Principles and Algorithm}
\noindent
We now have all the ingredients in place to describe {\sc iWarded}. As we have seen, its goal is enabling the generation of sets of Vadalog (i.e., Warded Datalog$^\pm$) rules, with diverse characteristics, to sustain tailored benchmarking for logic-based reasoning. In this section we delve into the core algorithm and the use case it serves. In particular, in Section~\ref{iWardedRules} we discuss the types of rules and features that can be generated, while Section~\ref{iWardedAlgorithm} is devoted to the core algorithm, whose correctness and theoretical challenges are discussed in Section~\ref{iWardedCorrectness}.
\label{iWarded}
\subsection{Generated Rules and Features}
\noindent
Let us illustrate {\sc iWarded} by starting from its output and characterize the types of rules and recursion it generates; we also touch on a set of additional features, part of the Vadalog language.

\smallskip
\noindent
\textbf{Types of Rules}.
{\sc iWarded} is able to generate \textit{existential} rules and, in particular, \textit{harmless}, \textit{harmful} and \textit{dangerous} rules. With regard to the structure of the rule itself, our generator can build \textit{linear rules} and \textit{join rules}: the former contain a single body atom, whereas in the latter multiple body atoms appear, whose facts are joined on the common variables; in the absence of shared variables, Cartesian product is implied.
As a specific design choice and without loss of generality, we do not generate join rules with more than two body atoms, as any n-ary join can be rewritten in this form~\cite{bellomarini2020vadalog}. Moreover, we consider only equi-joins on single variables. In this setting,
joins can be distinguished into \textit{harmless-harmless},
\textit{harmless-harmful} and \textit{harmful-harmful} depending on the nature of the involved variables.
The consequences of the presence of harmful-harmful joins in a set of rules and the respective simplification algorithm we propose will be extensively discussed in Section~\ref{hje}.
Harmless-harmless joins can in turn be of two sub-types: \textit{with ward}, when one of the involved atoms is a ward (and thus contains the dangerous variables), and \textit{without ward} otherwise (and thus there is no dangerous variable, by wardedness). Clearly, by wardedness, harmful-harmful joins do not contain dangerous variables.

\smallskip
\noindent
\textbf{Types of Recursion}. {\sc iWarded} can generate directly and indirectly recursive rules and, in the case of joins, rules with left, right and non-linear recursion. Let us define these categories.
We consider the \textit{predicate graph} of a set of rules $\Sigma$, built with a node labelled after each atom and a directed edge from node $\phi$ to $\psi$ if $\phi$ and $\psi$ respectively appear as a body atom and the head of some rule $\rho$ of $\Sigma$. The set $\Sigma$ is \textit{recursive} if the predicate graph is cyclic. Moreover, an atom is recursive if its corresponding node in the predicate graph is involved in a cycle and, finally, a rule is recursive if it contains at least one recursive atom.
A rule can be distinguished into \textit{directly recursive}, when a body atom also appears as its head, or \textit{indirectly recursive}, for longer cycles in the predicate graph.
For instance, the set of rules in Example~\ref{ex:mother} is recursive, and, in particular, the involved rules and atoms are all (indirectly) recursive.
With regards to join rules, different types of recursive rules are indeed possible, depending on whether one atom or both are recursive. We then talk about \textit{left}, \textit{right} and \textit{non-linear} recursive join rules, respectively.

\smallskip
\noindent
\textbf{Vadalog Expressions}. {\sc iWarded} supports many features and extensions of Vadalog, which cover much more than the Warded Datalog$^\pm$ core and are of practical utility. Here we consider and mention \textit{expressions}. They are 
Boolean expressions combining body variables (and ground values) with the standard comparison operators ($=$, $>$, $<$, $>=$, $<=$, $<>$), applicable to all data types. The semantics of an expression is that of an applicability constraint (or selection, or filter), conditioning the binding and thus the application of a rule. Multiple comma-separated conditions are allowed.
\label{iWardedRules}
\subsection{The Generation Algorithm}
\noindent
With the goal of testing the performance and evaluating the theoretical underpinnings of the Vadalog language, {\sc iWarded} can generate sets of rules with various characteristics and recursion structures, as we have seen in Section~\ref{iWardedRules}. At the same time, it guarantees high \textit{reusability} and \textit{user-friendliness}.
Reusability implies that the rules generated by {\sc iWarded} are not tightly coupled to a specific set of rules or scenario, but can be easily transported into and adapted for other scenarios; user-friendliness refers to the ergonomic design of {\sc iWarded}, which allows the user to be in full control of the definition of the set of rules to be generated, their characteristics and features, yet with limited amount of manual work and non-intrusive support for rule debugging and customization.

\begin{figure}[htb!]
  \centering
  \includegraphics[width=0.45\textwidth]{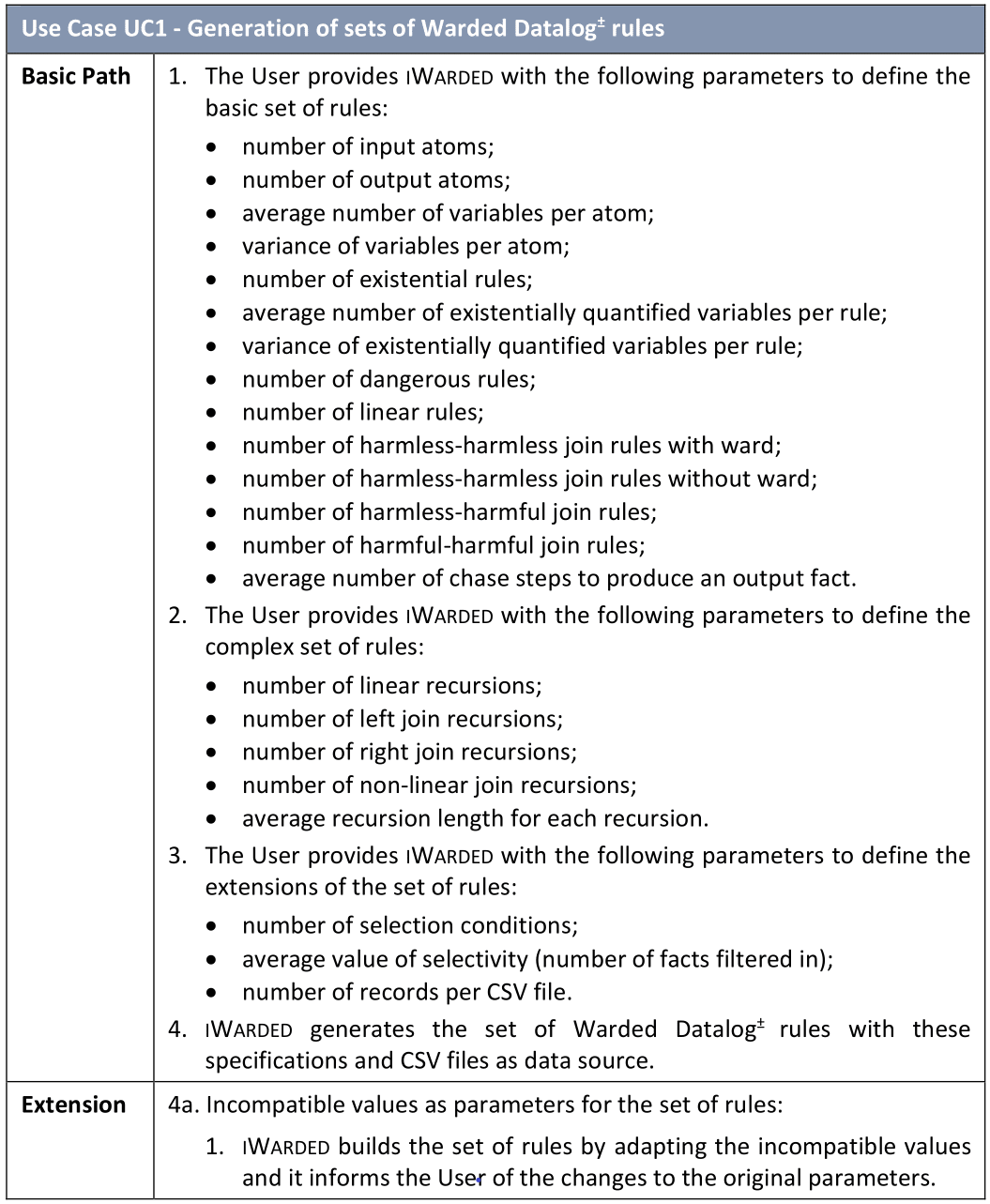}
  \caption{Main use case scenario of \textsc{iWarded}.}
  \label{fig:cases}\smallskip\smallskip
\end{figure}

\smallskip
\noindent
\textbf{\textsc{iWarded} Parameters}. An effective tactic to show the interaction of a user with {\sc iWarded} and therefore to highlight the involved parameters is by means of the \textit{main use case scenario}, as usual in formal software engineering~\cite{Larm04}. In Figure~\ref{fig:cases}, the basic path outlines the main parameters, required to generate the set of rules. Clearly, the values of the parameters have internal consistency conditions which are verified by {\sc iWarded} and, if needed, automatically adjusted and balanced for high tool usability.

\smallskip\smallskip\smallskip\smallskip

\noindent
\textbf{Generation Algorithm}. Algorithm~\ref{alg:iWardedGen} (\textit{iWardedGen}) takes as input the set of values for the parameters listed in Figure~\ref{fig:cases} and produces a set $\Sigma$ of Vadalog rules. The main idea of {\sc iWarded} here consists of abstracting any set of rules $\Sigma$ as a network of \textit{rule sequences}. A sequence is a chain of rules $\rho_1,\ldots,\rho_n$, where the head of $\rho_{i-1}$ appears as a body atom of $\rho_{i}$. A sequence can depart from another sequence or start from an input atom; likewise, a sequence can provide facts for another sequence or produce output facts. In another perspective, {\sc iWarded} controls the structure of $\Sigma$ by acting on the form of all the simple non-overlapping paths of its predicate graph, each corresponding to a sequence. The desiderata expressed via the parameters in Figure~\ref{fig:cases} are satisfied by generating suitable sequences (e.g., with a given number of atoms, variables, rules, existential quantifications, etc.) and globally balancing them.

Sequences of rules can be distinguished into \textit{input-output}, consisting of a chain where $\rho_n$ is an output atom and \textit{recursive}, when $\rho_n$ produces input facts for some rule $\rho_i$ with $0 \le i \le n$ of the sequence, closing the recursion. The length of input-output sequence depends on the \textit{average number of chase step} parameter; the average length of recursions is controlled by a dedicated parameter.

The choice of abstracting $\Sigma$ into a network of sequences provides a simple template mechanism to control the propagation of nulls and avoid violations of wardedness. Moreover, it produces standard sets of rules, easy to inspect, debug, adapt, reuse and communicate.

The algorithm (Algorithm~\ref{alg:iWardedGen}) proceeds as follows. After initializing an empty set of rules $\Sigma$ (step~1) with one single \textit{root atom}, $\Sigma$ undergoes an augmentation iteration in order to incrementally satisfy the requirements (with possible adjustments -- step~2) expressed with the input parameters. First, all the requirements for input-output and recursive sequences are satisfied by adding new linear rules and connecting them via join rules (steps~3-5). Then, each sequence is closed by either introducing recursion (step~6) or an output atom (step~8). To guarantee the global satisfaction of all the parameters, further off-sequence rules can be added (step~7). We next illustrate the overall structure of {\sc iWarded} output and see a real example, while for a more in-depth explanation of the algorithm steps, the reader is referred to the Appendix~\ref{appendixiWarded}.

\begin{algorithm}
\footnotesize
\caption{\texttt{iWardedGen}($\mathcal{S}$)}
\label{alg:iWardedGen}
\begin{algorithmic}[1]
\State {Initialize empty set of Vadalog rules $\Sigma$ from input scenario $\mathcal{S}$} \newline
\textcolor{darkblue}{\{Check compatibility of parameters and adapt if needed\}}
\State {\Call {\textcolor{darkgreen}{AdaptParametersCompatibility}}{$\mathcal{S}$}} \newline
\textcolor{darkblue}{\{Generate attributes for atoms, rules and sequences\}}
\State {\Call {\textcolor{darkgreen}{DefineAtomsRulesSequences}}{$\Sigma,\mathcal{S}$}} \newline
\textcolor{darkblue}{\{Create sequences of rules for chase steps of output atoms\}}
\State {\Call {\textcolor{darkgreen}{GenerateRulesInputOutputSequence}}{$\Sigma,\mathcal{S}$}} \newline
\textcolor{darkblue}{\{Create sequences of rules based on length of recursion\}}
\State {\Call {\textcolor{darkgreen}{GenerateRulesRecursiveSequence}}{$\Sigma,\mathcal{S}$}} \newline
\textcolor{darkblue}{\{Create rules for direct and indirect recursive closure\}}
\State {\Call {\textcolor{darkgreen}{GenerateRulesRecursiveClosure}}{$\Sigma,\mathcal{S}$}}\newline
\textcolor{darkblue}{\{Create rules for remaining input requirements\}}
\State {\Call {\textcolor{darkgreen}{GenerateRulesParameterCompletion}}{$\Sigma,\mathcal{S}$}} \newline
\textcolor{darkblue}{\{Create rules with output atoms as heads\}}
\State {\Call {\textcolor{darkgreen}{GenerateRulesOutputClosure}}{$\Sigma,\mathcal{S}$}}
\end{algorithmic}
\end{algorithm}

\begin{figure}[hbt!]
  \centering
  \includegraphics[width=0.45\textwidth]{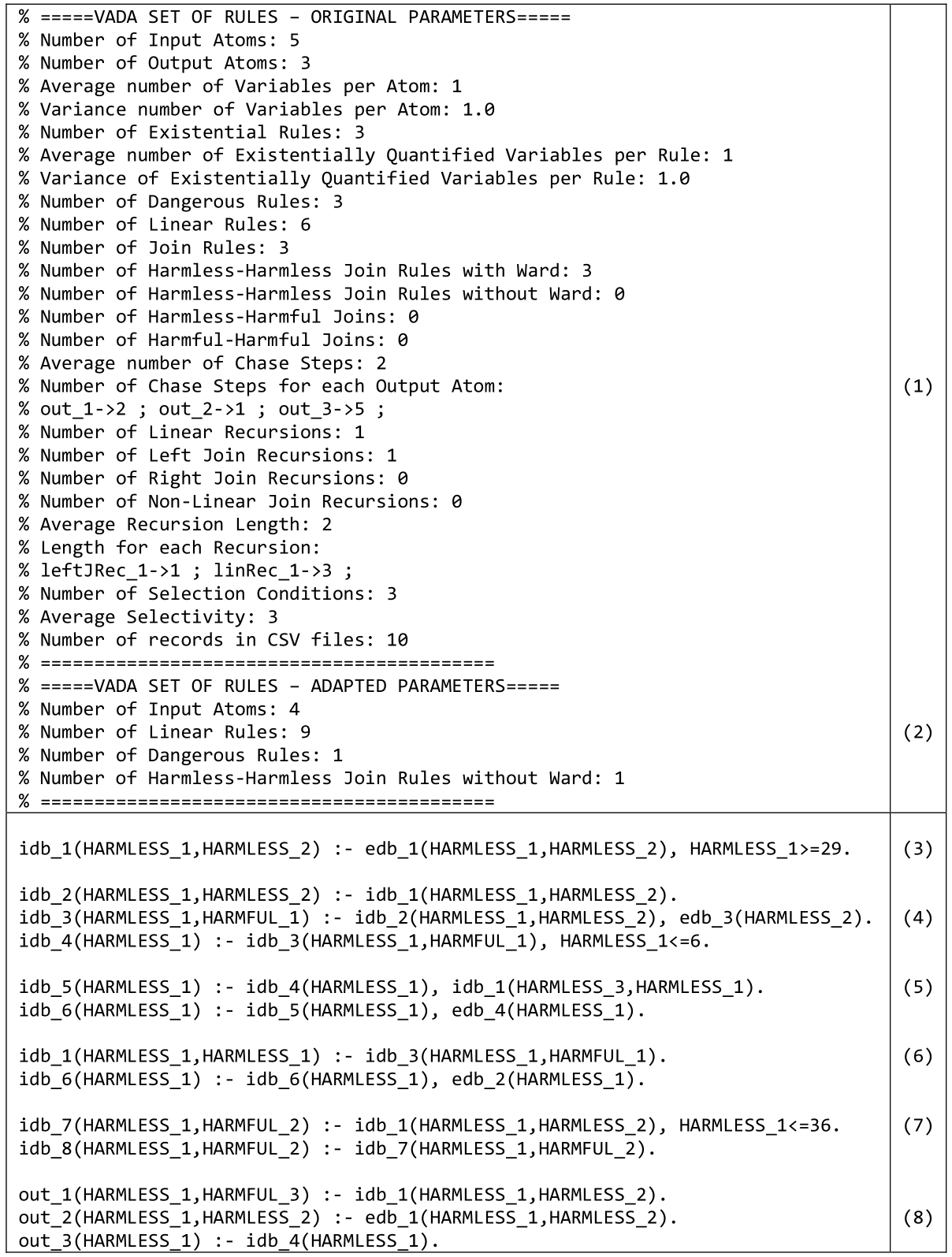}
  \caption{Example of generated set of rules.}
  \label{fig:example}
\end{figure}

\noindent
\textbf{Generated Set of Rules}. Consider the set of rules in Figure~\ref{fig:example}, which shows the output of a run with {\sc iWarded}.

\begin{enumerate}[noitemsep,nolistsep,leftmargin=2mm]
    \item \textit{Comment section - Original parameters:} documents the value of the input parameters as well as internal parameters generated by {\sc iWarded} (e.g., the average number of chase steps and the number of steps for each output atom, respectively).
    \item \textit{Comment section - Adapted parameters:} documents the adjustments applied by {\sc iWarded} to the input parameters to generate a consistent set of rules.
    \item \textit{Rule with root atom:} the first rule of the set, by convention it is a linear harmless rule.
    \item \textit{Input-Output sequence:} a set of linear and join rules, with various characteristics based on the input parameters.
    \item \textit{Recursive sequences:} a set of rules part of a recursion.
    \item \textit{Recursive closure sequence:} a set of linear and join rules, each introducing a recursion and thus a cycle in the predicate graph.
    \item \textit{Parameter completion sequence:} a set of extra rules needed to globally satisfy the parameters, e.g., the number of selection conditions, in the form of expressions, in our example.
    \item \textit{Output completion sequence:} a set of linear rules, one for each input-output sequence; the rule head is the output atom and the body atom is connected to the head of the rule closing the sequence.
\end{enumerate}

\smallskip\smallskip\smallskip\smallskip\smallskip

\noindent
\textbf{Data Generation}. For space reasons we omit here to discuss the data generation mechanisms adopted by {\sc iWarded}. In brief, we produce CSV files (or DB tuples) with data having various distributions according to the input parameters, e.g., to induce specific selectivity, guarantee the applicability of joins, etc. Such files are bound to input atoms with an annotation-based mapping mechanism.
\label{iWardedAlgorithm}
\subsection{Algorithm Discussion}
\noindent
Ensuring the wardedness of the generated set of rules $\Sigma$ is a central correctness guarantee that Algorithm~\ref{alg:iWardedGen} must fulfil and the challenge is by no means trivial. 
In fact, {\sc iWarded} is tasked to balance the need to generate a given number of rules of different types and with diverse features (e.g., having existentials, harmless and harmful rules, join rules, etc.) combined into sequences while guaranteeing wardedness of $\Sigma$. The na\"ive brute-force approach would consist in simply checking the wardedness condition we have seen in Section~\ref{vadalog} after the execution of each step, amending the generated rules by backtracking when needed, until convergence. This would be computationally very costly, with possible exponential blowup in the number of rules. Thanks to our sequence network structure, we proceed inductively and incrementally add rules to sequences in such a way that each addition does not hamper wardedness. A full proof by induction of the correctness of our algorithm is beyond the scope of this paper and possibly uninteresting; instead, we show a number of insightful examples, disclosing how wardedness is preserved in the most relevant cases when building rule sequences.

\begin{example}
\label{ex:harmlessJoin}
\begin{ttmath}
\begin{align*} 
1:idb_1(h_1,h_2) :- \,\,edb_1(h_1,h_2).\\
2:idb_2(h_2,{\overline h_3}) :- \,\,idb_1(h_1,h_2).\\
3:out_1(h_1,{\overline h_3}) :- \,\,idb_1(h_1,h_2),idb_2(h_2,{\overline h_3}).
\end{align*}
\end{ttmath}
This example shows an input-output sequence with \texttt{edb\textsubscript{1}} as input atom and \texttt{out\textsubscript{1}} as output atom. The variable $h_3$ in rule \texttt{2} is existentially quantified, thus the affected position (denoted by overline) is propagated to \texttt{idb\textsubscript{2}} in the body of rule \texttt{3}, which is a harmless-harmless join with ward. The head of rule \texttt{3} contains the harmful variable present in \texttt{idb\textsubscript{2}}, the ward, therefore the sequence is warded.
\end{example}

\begin{example}
\label{ex:harmfulJoin}
\begin{ttmath}
\begin{align*} 
1:idb_1(h_1,{\overline h_3}) :- \,\,edb_1(h_1,h_2).\\
2:idb_2(h_1,{\overline h_3}) :- \,\,idb_1(h_1,{\overline h_3}).\\
3:out_1(h_1,h_2) :- \,\,idb_1(h_1,{\overline h_3}),idb_2(h_2,{\overline h_3}).
\end{align*}
\end{ttmath}
This example shows an input-output sequence with \texttt{edb\textsubscript{1}} as input atom and \texttt{out\textsubscript{1}} as output atom. The second argument in the head of rule \texttt{1} is existentially quantified, therefore the affected position is propagated to atoms \texttt{idb\textsubscript{1}} and \texttt{idb\textsubscript{2}} in the body of rule \texttt{3}, which is a harmful-harmful join. The head of rule \texttt{3} is harmless, thus respecting the wardedness of the sequence.
\end{example}

\begin{example}
\label{ex:harmfulRec}
\begin{ttmath}
\begin{align*} 
1:idb_1(h_1,{\overline h_3}) :- \,\,edb_1(h_1,h_2).\\
2:idb_2(h_1,{\overline h_3}) :- \,\,idb_1(h_1,{\overline h_2}).\\
3:idb_3(h_1,{\overline h_3}) :- \,\,idb_1(h_1,{\overline h_2}).\\
4:idb_1(h_1,{\overline h_4}) :- \,\,idb_2(h_1,{\overline h_3}),idb_3(h_2,{\overline h_3}).
\end{align*}
\end{ttmath}
This example shows two indirect recursive sequences. The second term in the head of rules \texttt{1}, \texttt{2} and \texttt{3} is existentially quantified, therefore the affected position is propagated to atoms \texttt{idb\textsubscript{2}} and \texttt{idb\textsubscript{3}} in the body of rule \texttt{4}, which is a harmful-harmful join. The head of rule \texttt{4} is \texttt{idb\textsubscript{1}}, which closes the recursion. It contains a new existential $h_4$ and does not propagate $h_3$. Therefore, the sequence is warded.
\end{example}

\noindent
The examples provide insight into the rationale of iWardedGen for the propagation of affected positions.
Example~\ref{ex:harmlessJoin} and Example~\ref{ex:harmfulJoin} show two distinct decisions of the algorithm about how to extend the sequence with rule \texttt{3}: in the former, it is assumed that the input parameters require an harmless-harmless join rule with ward, therefore the join is between the second term of \texttt{idb\textsubscript{1}} and the first one of \texttt{idb\textsubscript{2}} (the ward), both harmless; in the latter, the join rule is required to be harmful, therefore a join is added between the second harmful terms of the two atoms involved.

\smallskip
In iWardedGen we adopt a \textit{memory-free generation} approach. While the generation of rules orderly proceeds, the affected positions on a rule only depend on the rules that have been previously generated. In other terms, recursive rules do not introduce new affected positions. Of course, this does not prevent them from propagating existentially quantified variables or having existential quantification. This is apparent in Example~\ref{ex:harmfulRec}, where $h_4$ is existentially quantified in rule~4, but position $\texttt{idb}_1[2]$ is already affected because of existential quantification in rule~1. Memory-free generation has many advantages: first, no complex memory structures are needed to keep track of propagation of nulls; wardedness is easy to enforce, verify and visually check by inspecting even lengthy rule sequences; once a rule has been classified as warded, the property is an invariant throughout the algorithm execution, and need not be revised, a valuable property for overall correctness check.
\label{iWardedCorrectness}
\section{Vadalog Normalizer}
\noindent
As we have introduced, in order to exploit reasoning boundedness results, Warded Datalog$^\pm$ scenarios must not contain harmful joins, and thus expressed in \textit{Harmless Warded Datalog$^\pm$}. In this case, although $\Sigma(D)$ (Section~\ref{vadalog}) is potentially infinite because of the generation of infinite labelled nulls, for the reasoning task, the {\sc chase} can be considered up to isomorphism of facts, and so aggressively pruned. Intuitively, the absence of joins on labelled nulls makes their identity irrelevant in isomorphism evaluation, as fully detailed in~\cite{bellomarini2018vadalog}. This leads to practical algorithms guaranteeing termination and low memory footprint. A specific strategy for harmful join elimination is therefore central to build benchmarks for Warded Datalog$^\pm$ that have \textit{full coverage}, in that all the language features (including harmful joins) can be used, and are \textit{fair}, in the sense that decouple normalization time from reasoning evaluation. 
It, perhaps, may also lead to different normalization techniques becoming of interest to Datalog$^\pm$ reasoning systems to exploit the wardedness.

\smallskip
This section thus presents the \textbf{Vadalog normalizer}, with the  \textbf{Harmful Join Elimination} (HJE) algorithm at its center.
At all times, such an algorithm must:
\begin{itemize}[noitemsep,topsep=0pt,leftmargin=2mm]
    \item respect equivalence between the original set of rules and the normalized one, thus avoiding any loss of meaning;
    \item allow to operationally guarantee both termination and correctness of the reasoning tasks.
\end{itemize}

\smallskip
\noindent
In Section~\ref{hjePreliminaries} we present some relevant preliminary concepts. In Section~\ref{hjeTheoreticalResult} we illustrate and prove the theory behind the normalization of sets of rules. In Section~\ref{hjeAlgorithm} we provide an in-depth analysis of the algorithm designed. In Section~\ref{hjeCorrectness} we show the correctness of our approach and discuss additional properties of interest.
\label{hje}
\subsection{Preliminary Concepts}
\noindent
We now present the preliminary notions at the basis of our normalization algorithm for the fragment Warded Datalog$^\pm$.

\noindent
\textbf{Homomorphisms and Isomorphisms}. An atom \texttt{A} is \textit{homomorphic} to an atom \texttt{B} if there exists a substitution $\theta$ for variables of \texttt{A} such that $\theta\texttt{A} = \texttt{B}$. If there exists a bijective $\theta$, then \texttt{A} and \texttt{B} are \textit{isomorphic}. Moreover, a rule $\rho$ presents a \textit{(partial) body-homomorphism} with a rule $\sigma$ if there exists a homomorphism between (some of) the atoms in the respective bodies: as said above, if bijective, it is called \textit{(partial) body-isomorphism}. 

\noindent
\textbf{Dom Atoms}. \textit{dom(*)} is an artificial body atom ensuring that all  variables in the body bind only against ground values in the domain. It is used to avoid the propagation of labelled nulls.

\noindent
\textbf{Skolem Terms}. A \textit{Skolem function} calculates the values for existentially quantified variables, to control the identity of labelled nulls. It is injective, deterministic and range disjoint. A \textit{Skolem term} has the form $f_{iy}(y_1,...,y_n)$ and expresses that $y_1,...,y_n$ are operands of a Skolem function $f_i$ for the existentially quantified variable \textit{y}.

\noindent
\textbf{Existential Unfolding}. Let $\rho$ be a rule \texttt{A},\texttt{K}$\to$\texttt{B}, where \texttt{A} and \texttt{B} are atoms and \texttt{K} a conjunction of atoms, and $\sigma$ be a rule \texttt{D}$\to$\texttt{A$^\prime$}, whose head is homomorphic to the atom \texttt{A} in $\rho$ (that is, there exists a substitution $\theta$ such that $\theta\texttt{A} = \texttt{A}^\prime$). The result of \textit{unfolding} $\rho$ at \texttt{A} is the rule {$\tau$} (\texttt{D},\texttt{K}$\to$\texttt{B})$\theta$ and we define $\rho$ as the \textit{ancestor} of {$\tau$}. Moreover, we extend the unfolding procedure to the case in which the homomorphic head contains an existentially quantified variable \textit{y}: the result of \textit{existential unfolding} is the one obtained from regular unfolding but with a Skolem atom $f_{iy}$ for the variable \textit{y}.

\noindent
\textbf{Folding}. Let $\rho$ be a rule \texttt{B}$\to$\texttt{H}, where \texttt{H} is an atom and \texttt{B} an atom or a conjunction of atoms, and $\sigma$ be a rule such that there is (full or partial) body-isomorphism with $\rho$ by substitution $\theta$. The result of \textit{folding} $\rho$ into $\sigma$ is the rule {$\tau$} (\texttt{B$^\prime$}$\to$\texttt{H})$\theta$ where \texttt{B$^\prime$} is the head of \textit{$\sigma$} joined with the atoms in \texttt{B} which were not part of the body-isomorphism.

\noindent
\textbf{Causes of Affectedness}. Let $\Sigma$ be a set of rules and $\rho$ a \textit{harmful join rule} (a rule containing a harmful join in the body). As stated in Section~\ref{iWardedRules}, such rules can always be rewritten (by breaking more complex rules into multiple steps) as having exactly one harmful join, in the following form: 
\begin{center}
    $\rho:\forall x\forall y\forall{\overline h}(A(x_1,y_1,{\overline h}),B(x_2,y_2,{\overline h})\to\exists z\,C(x,z))$.
\end{center}
By definition of harmful variables, $\Sigma$ will also contain (at least) one set of rules $\Gamma_I=\{\sigma_1,...\sigma_m\}$ for every distinct atom \texttt{I} with a position involved in the harmful join, each containing:
\begin{itemize}[noitemsep,nolistsep,leftmargin=2mm]
    \item one \textit{direct} cause of affectedness, that is an existential rule which causes a position to be affected, in the following form:
    \begin{center}
        $\sigma_1:\forall x\forall y(F(x,y)\to\exists{h}\,E(x,y,{\overline h}))$
    \end{center}
    \item zero (or more) \textit{indirect} causes of affectedness, that are rules propagating an affected position from the body to the head, as follows:
    \begin{center}
        $\sigma_2:\forall x\forall y\forall {\overline h}(E(x,y,{\overline h})\to\,D(x,y,{\overline h}))$ \\
        $\sigma_3:\forall x\forall y\forall {\overline h}(D(x,y,{\overline h})\to\,A(x,y,{\overline h}))$
    \end{center}
\end{itemize}
such that the head of the previous cause of affectedness is in the body of the next one (from the direct one to the last indirect one, which propagates the affected position to $\rho$ itself). The rules above are causes for the affected position in atom \texttt{A} involved in the harmful join ($\Gamma_{A}$); as previously stated, $\Sigma$ will also contain a corresponding set of causes for \texttt{B} ($\Gamma_{B}$).
\label{hjePreliminaries}
\subsection{Normalization Problem}
\noindent
We formulate the problem of normalizing a set of rules as the removing of harmful join rules. Let $\Sigma$ be a Warded Datalog$^\pm$ 
set of rules with one or more harmful join rules. The \textit{normalization problem} consists in finding an equivalent (i.e., meaning-preserving~\cite{afrati2003linearisability}) Harmless Warded Datalog$^\pm$
set of rules $\Sigma^\prime$.
We argue that the problem above is always solvable by proving the following theorem.

\begin{theorem}
\label{th:hjeharmlesswarded}
For any Warded Datalog$^\pm$ set of rules $\Sigma$ with harmful join rules there exists an equivalent (i.e., meaning-preserving) Harmless Warded Datalog$^\pm$ set of rules $\Sigma^\prime$.
\end{theorem}

\noindent
With reference to the preliminary concepts described above, we define a support structure at the basis of our proof. 
\begin{definition}
Let $\Sigma$ be a set of rules, $\rho$ a rule and $\Phi$ a (partial) function from rules to atoms. An \textbf{unfolding tree (U-tree)} \textit{T} for \textit{$\big\langle\Sigma$,$\rho\big\rangle$} is a tree labelled with rules, constructed as follows~\cite{afrati2003linearisability}:
\begin{itemize}[noitemsep,nolistsep]
	\item $\rho$ is the root label of \textit{T};
	\item if \textit{M} is a node labelled by a rule $\nu$ and \texttt{I} is the atom selected by $\Phi$ in $\nu$, then, for each rule $\nu^\prime$ in the result of unfolding $\nu$ at \texttt{I}, there is a child node \textit{N} of \textit{M} labelled by $\nu^\prime$.
\end{itemize}
Let $\Sigma$ be a set of rules, $\rho$ a harmful join rule and $\Gamma_A,\Gamma_B$ the sets of causes of affectedness for $\rho$ in $\Sigma$. We define \textbf{harmful unfolding tree (HU-tree)} \textit{T} for \textit{$\big\langle\Sigma$,$\rho\big\rangle$} as a U-tree such that $\Phi$ is a function which selects the atom \texttt{I} in $\rho$ involved in the harmful join and each rule $\rho^\prime$ is the result of unfolding $\rho$ at \texttt{I} with a cause of affectedness in $\Gamma_I$ ($\rho$ is the \textit{ancestor} of $\rho^\prime$).
\end{definition}

\noindent
\begin{definition}
Let $\Sigma$ be a set of rules, let $\rho$ be a harmful join rule\\
$\rho: A,B \to C$ in $\Sigma$ and let $\Gamma$ be a multiset which contains the causes of affectedness for $\rho$ in \texttt{A} and \texttt{B} from a pair of sets $\Gamma_A,\Gamma_B$ (respectively). We define the \textbf{distance from harmlessness (\texttt{dh})} of $\rho$ in $\Sigma$ as the cardinality of $\Gamma$. Let the \textbf{maximum distance from harmlessness (\texttt{mdh})} of $\rho$ in $\Sigma$ be the maximum of the \texttt{dh} for $\rho$.
\end{definition}

\begin{figure}[hbt!]
  \centering
  \includegraphics[width=0.33\textwidth]{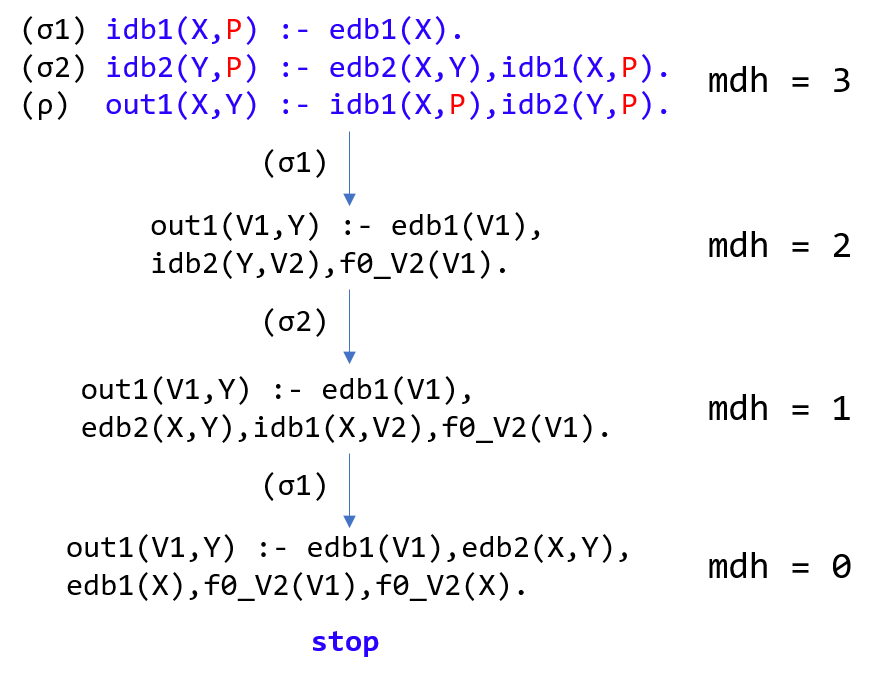}
  \caption{Example of branch in basic HU-Tree.}
  \label{fig:hutree}
\end{figure}

\noindent
Figure~\ref{fig:hutree} shows a basic example of a branch (a root-to-leaf path) in a HU-Tree, derived from unfolding the harmful join rule $\rho$ with its causes of affectedness $\sigma_1$ and $\sigma_2$ at \texttt{idb1} and \texttt{idb2}, respectively. Here the only pair of causes of affectedness is $\Gamma_{idb1}=\{\sigma_1\}$ and $\Gamma_{idb2}=\{\sigma_1,\sigma_2\}$, therefore $\Gamma=[\sigma_1,\sigma_1,\sigma_2]$ and \texttt{dh}=\texttt{mdh}=3.

\begin{proof}[Proof (of Theorem~\ref{th:hjeharmlesswarded})] Let $\Sigma$ be a Warded Datalog$^\pm$ set of rules with $n$ harmful join rules $\rho_1,...,\rho_n$. We prove Theorem~\ref{th:hjeharmlesswarded} by showing that the following statements hold:
\begin{enumerate}[noitemsep,nolistsep]
	\item \label{itm:termination} there exists a procedure to build the HU-Tree $T_i$ for $\rho_i$ that always terminates;
	\item \label{itm:correctness} the set of logic rules $\Sigma^\prime$ obtained by such a procedure is a Harmless Warded Datalog$^\pm$ set of rules 
	equivalent to $\Sigma$ in the sense it is meaning-preserving.
\end{enumerate}

\smallskip
(\ref{itm:termination}). Let $\rho_i$ be the $i$-th harmful join rule in $\Sigma$ and $\Gamma$ be the multiset with the longest combination of causes, i.e., \texttt{dh}=\texttt{mdh}. We define the HU-Tree $T_i$ with $\rho_i$ as root and proceed by induction on \texttt{mdh}.

\noindent
\textbf{Base case (\texttt{mdh}=1)}: by definition of causes of affectedness, $\Gamma$ only contains the direct cause $\sigma_1$. By applying \textit{existential unfolding} to $\rho_i$ with $\sigma_1$ we obtain a rule without causes of affectedness, a \textit{leaf} of $T_i$. Thus \texttt{mdh}=\texttt{mdh}-1=0 and the corresponding branch is closed.\\
\textbf{Inductive hypothesis (\texttt{mdh}=h)}: let us assume the procedure to build the current branch in $T_i$ terminates.

\noindent
\textbf{Induction step (\texttt{mdh}=h+1)}: by definition of causes of affectedness, $\Gamma$ contains a sequence $\sigma_2,...,\sigma_{h}$ of $h-1$ indirect causes, and two direct causes $\sigma_1$. 

Now, by applying \textit{unfolding} to $\rho_i$ with $\sigma_{h}$ we obtain a new rule $\rho_i^\prime$. We attempt \textit{folding} $\rho_i^\prime$ with $\rho_i$ (currently, its only ancestor in the branch): if it succeeds, by definition of folding and by construction of $T_i$ we obtain a rule without causes of affectedness (a \textit{leaf}), thus \texttt{mdh}=0 and the corresponding branch is closed; otherwise, we remove $\sigma_{h}$ from $\Gamma$ and so the corresponding \texttt{mdh}=\texttt{mdh}-1=$h$. Hence, by inductive hypothesis the procedure to build the current branch in $T_i$ terminates. Finally, we observe that, by definition of \textit{maximum distance from harmlessness} and by construction of the HU-Tree, all the other branches in $T_i$ are closed at the same time or before the one above. Therefore the procedure always terminates in a finite number of steps.

\smallskip
(\ref{itm:correctness}). Let $T_1,...,T_n$ be the HU-Trees created with the described procedure and let $\Sigma^\prime$ be a new set of rules built from $\Sigma$ as follows:
\begin{enumerate}[noitemsep,nolistsep,label=(\roman*)]
	\item remove $\rho_1,...,\rho_n$ from $\Sigma$;
	\item add the \textit{leaves} of $T_1,...,T_n$ to $\Sigma$;
	\item for each $\rho_i$, add the \textit{grounding} rules below which operate only on EDB facts (that is, imported from an external data source, such as a relational database) using the \textit{Dom} feature, to avoid the propagation of nulls.\\
	Dom$({\overline h}),A(x_1,y_1,{\overline h}) \to A^\prime(x_1,y_1,{\overline h})$\\
    $A^\prime(x_1,y_1,{\overline h}) \to A(x_1,y_1,{\overline h})$\\
    $A^\prime(x_1,y_1,{\overline h}),B(x_2,y_2,{\overline h})\to\exists z\,C(x,z))$
\end{enumerate}
Now we observe that:
\begin{itemize}[noitemsep,nolistsep,leftmargin=2mm]
	\item $\Sigma^\prime$ is \textbf{harmless}. This holds by construction, as the procedure terminates only when every $\rho_i$ has been unfolded with all the causes of affectedness or has been folded with one of its ancestors;
	\item $\Sigma^\prime$ is \textbf{warded}. This holds by definition of folding and unfolding, as they produce rules which do not add new existentials. Therefore, as $\Sigma$ is warded by hypothesis, $\Sigma^\prime$ is warded as well;
	\item $\Sigma^\prime$ is \textbf{equivalent} to $\Sigma$. 
	Two sets of logic rules are considered equivalent if they have the same meaning~\cite{afrati2003linearisability}. To define such meaning, many semantics have been adopted in the literature since the early times~\cite{EmKo76}. As anticipated in Section~\ref{vadalog}, in this work we follow a practical approach and operationally define the meaning of a set of rule $\Sigma$ via the well-known {\sc chase} procedure~\cite{MaMS79}, so that $\Sigma$ and $\Sigma^\prime$ are equivalent if $\Sigma(D) = \Sigma^\prime(D)$ for every database instance $D$. In our setting, by construction, $\Sigma$ and $\Sigma^\prime$ only differ by the rules produced via grounding, folding and unfolding, for which {\sc chase}-equivalence can be easily derived for Datalog$^\pm$ as a generalization of proofs in the Datalog context~\cite{TaSa84}. \qedhere
	\end{itemize}
\end{proof}

\noindent
Given the above definitions and procedure to build a HU-Tree to normalize a generic Warded Datalog$^\pm$ set of rules, we now make the following observation regarding this structure. This will also allow us to discuss time bounds later.

\begin{proposition}
\label{prop:hjetreecomplexity}
Let $\Sigma$ be a Warded Datalog$^\pm$ set of rules with a harmful join rule $\rho$ s.t.\
\texttt{mdh}=$h$ and $s$ is the number of multisets $\Gamma$, and let $T$ be the corresponding HU-Tree with $\rho$ as root. We show that:
\begin{enumerate}[noitemsep,nolistsep]
    \item \label{itm:numnodes} the number of nodes in $T$ (rules created in the building of $T$) \texttt{numNodes} is in ${O}(s*h)$;
    \item \label{itm:numfolds} the number of folding checks in $T$ \texttt{numFolds} is in ${O}(s*h^2)$;
    \item \label{itm:numexp} \texttt{numNodes} and \texttt{numFolds} are polynomial in the branching factor $k$ (i.e., the number of unfolding for each node of $T$) and exponential in \texttt{mdh} for $\rho$ in $\Sigma$;
    \item \label{itm:numtrees} the procedure shows a linear behaviour in the number of harmful join rules.
\end{enumerate}
\end{proposition}

\begin{figure*}[!htb]
\centering
 \begin{tabular}{@{}cc@{}}
    \subfloat{%
    \includegraphics[width=.49\textwidth]{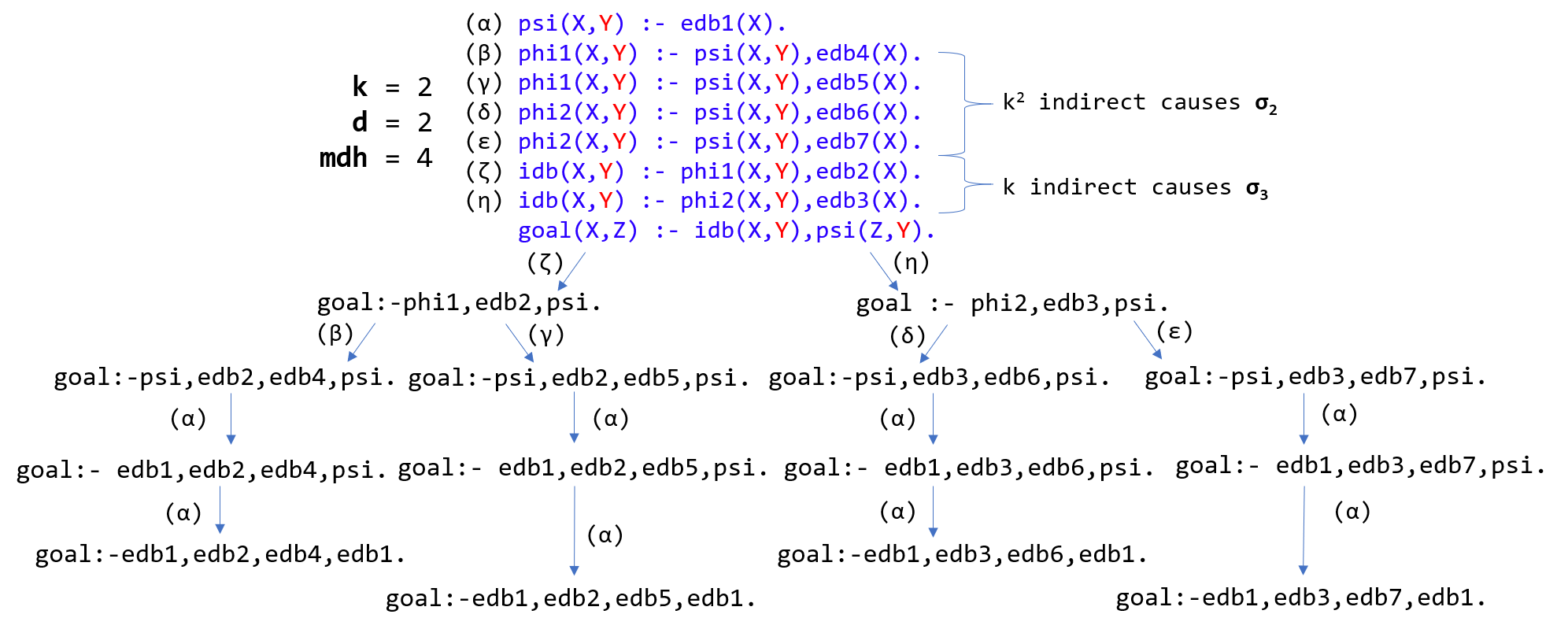}}
    \subfloat{%
    \includegraphics[width=.49\textwidth]{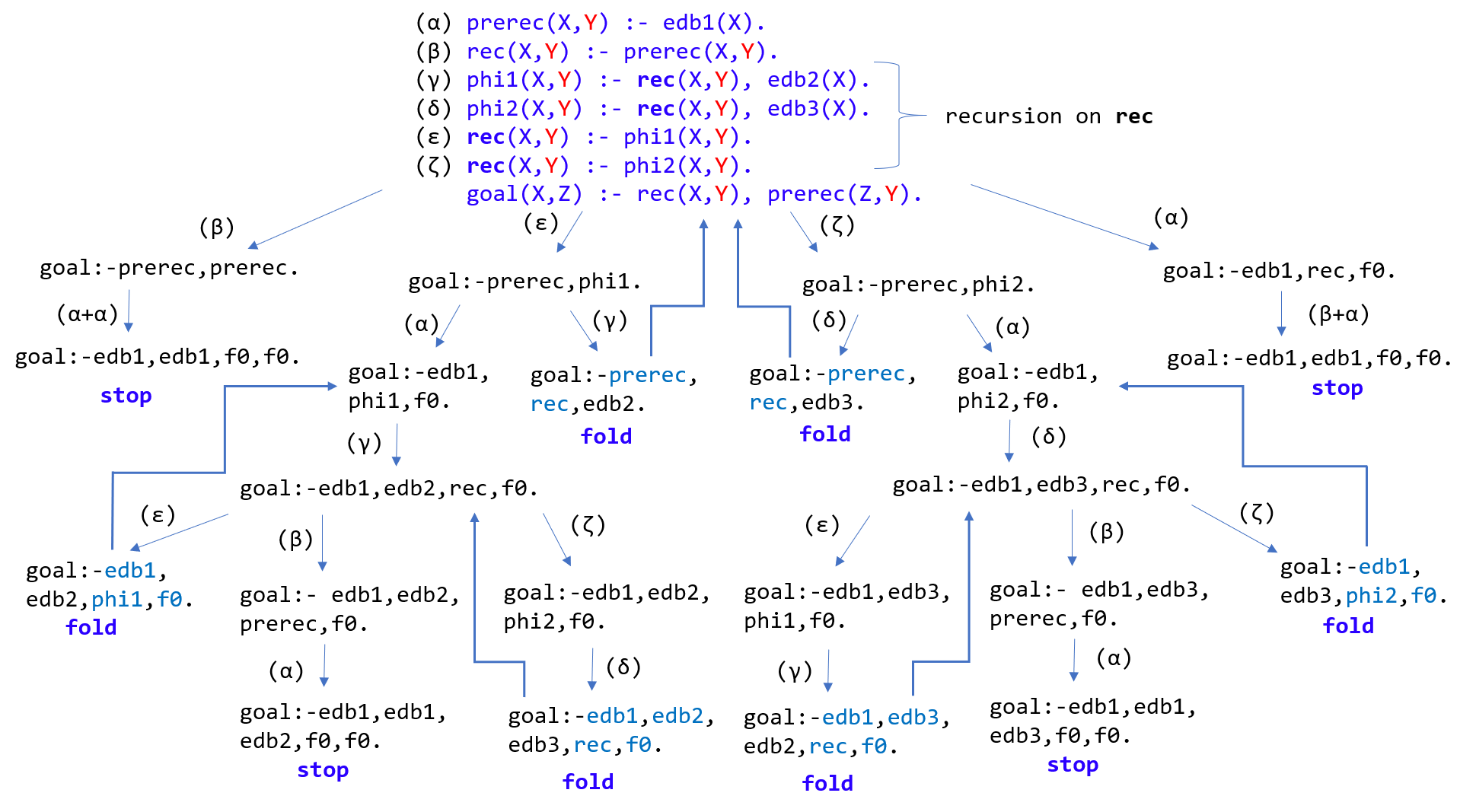}}
 \end{tabular}
 \caption{Relevant scenarios for HJE complexity (shown in large in the appendix).}
  \label{fig:hjerelevantcases}
  \smallskip\smallskip\smallskip\smallskip\bigskip
\end{figure*}

\begin{proof}
By definition of HU-Tree, \texttt{mdh} and $\Gamma$, $T$ consists of $s$ distinct branches with (at most) $h$ nodes each.

(\ref{itm:numnodes}). By construction, we have that \texttt{numNodes} $\in {O}(s*h)$.

(\ref{itm:numfolds}). By definition of folding in the procedure, every node attempts folding with each of its ancestors in the branch. In a single branch, at \texttt{mdh}=$h-1$ the node makes a single attempt with the only ancestor present, that is the root, whereas at \texttt{mdh}=0 the node attempts (up to) $h$ folding checks. Therefore, the total number of folding checks in the single branch is (at most) $h(h+1)/2$ and \texttt{numFolds} $\in {O}(s*h^2)$.

(\ref{itm:numexp}). Following the notation provided in Section~\ref{hjePreliminaries}, let $d$ be the length of each sequence of indirect causes of affectedness, from the one linked to the direct cause to the last indirect one which links to the harmful join rule (that is, \texttt{mdh}-2 as we are not considering the direct cause for both the atoms involved in the join), and let $\Sigma$ be a set of rules built as follows: $k$ rules $\sigma_d^\prime,...,\sigma_d^k$ whose head contains an atom in the body of $\rho$, and $k^2$ rules $\sigma_{d-1}^\prime,...,\sigma_{d-1}^{k^2}$ whose head contains the atom with affected position of the corresponding $\sigma_d^i$ and so on. By definition of $\Gamma$, the number of branches in $T$ is $s\in {O}(k^d)$. Therefore, by construction of HU-Tree and by (1) and (2), both \texttt{numNodes} and \texttt{numFolds} are polynomial in the
branching factor $k$ and exponential in \texttt{mdh}.
Figure~\ref{fig:hjerelevantcases}(a) shows a basic example of the scenario described above.

(\ref{itm:numtrees}). By definition, each harmful join rules causes the creation of a single HU-Tree, thus proving the linear behaviour.
\end{proof}

\begin{figure*}[!htb]
\centering
 \begin{tabular}{@{}cc@{}}
    \subfloat{%
    \includegraphics[width=.49\textwidth]{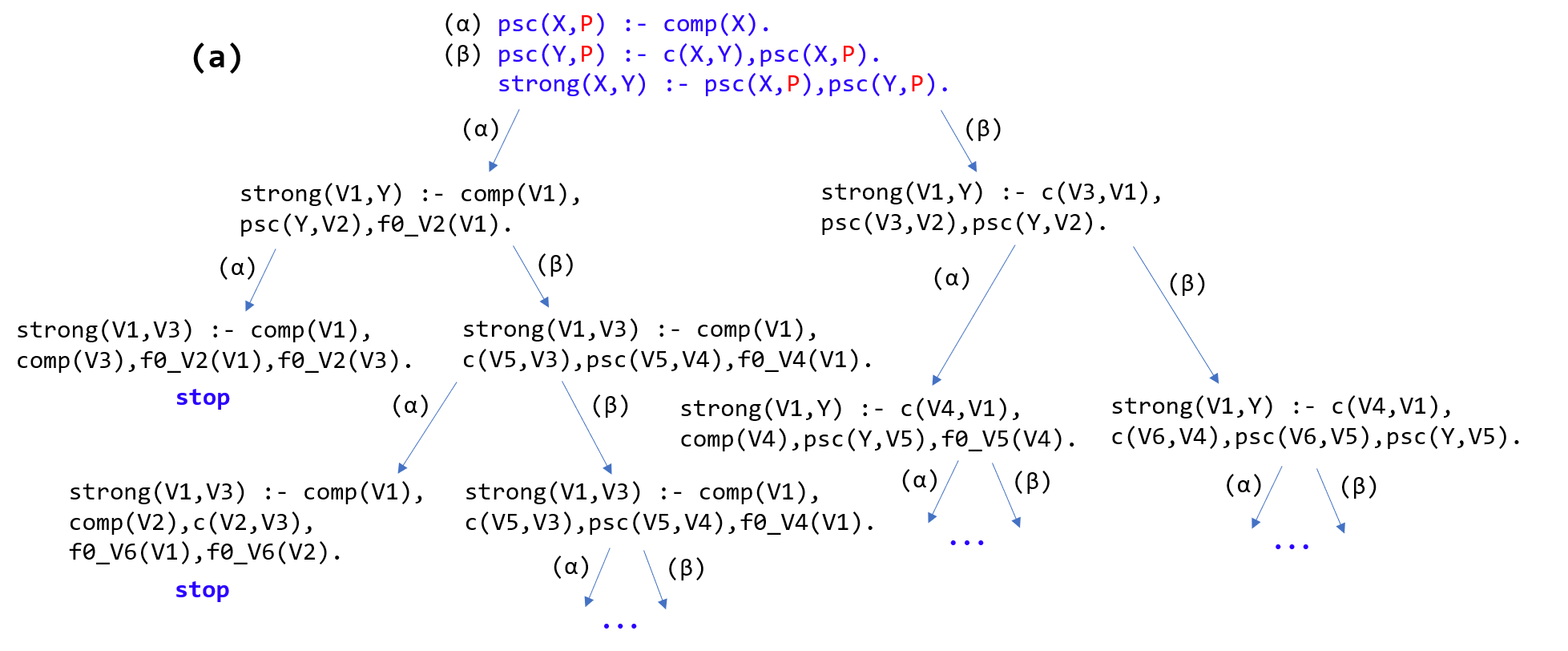}}\hfill
    \subfloat{%
    \includegraphics[width=.49\textwidth]{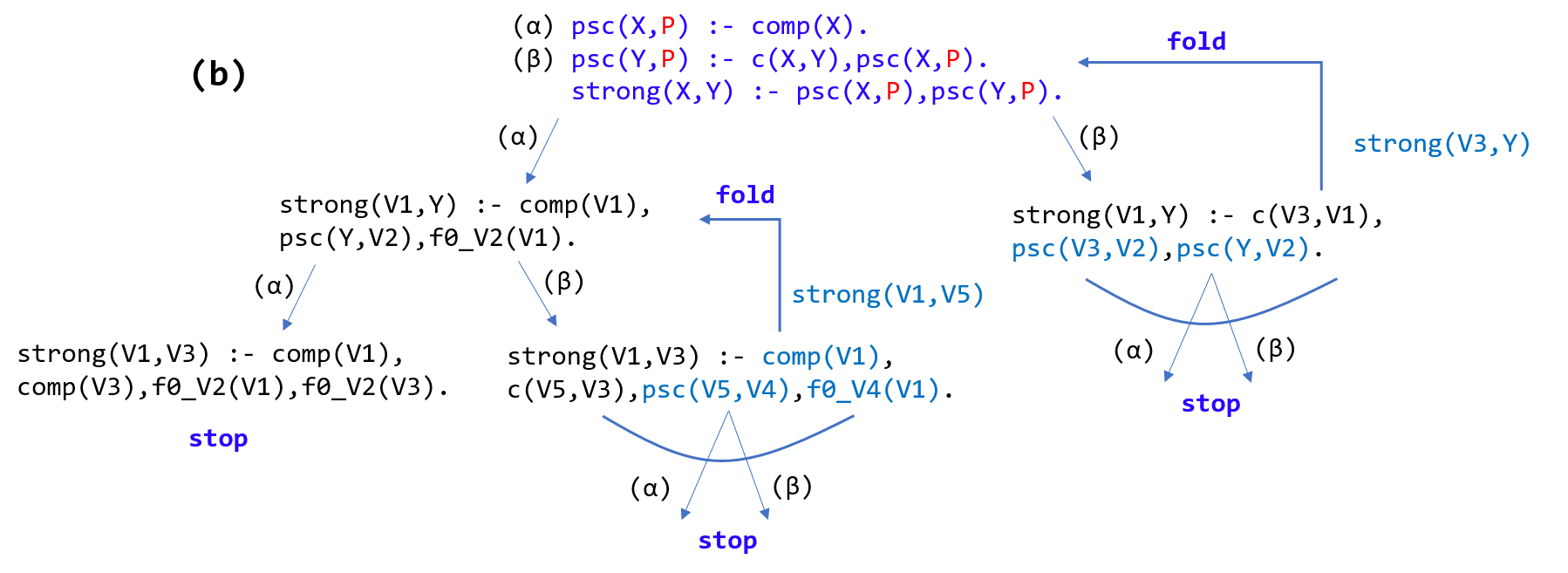}}\hfill
 \end{tabular}
 \caption{Role of Folding in HJE termination (shown in large in the appendix).}
  \label{fig:hjeexecution}
\end{figure*}\smallskip
\label{hjeTheoreticalResult}
\subsection{HJE Algorithm}
\noindent
The proof of Theorem~\ref{th:hjeharmlesswarded} about the existence of a harmless version of a given set of rules $\Sigma$ gives the inspiration for an algorithm workable in practice. In the following,
we provide the algorithm designed to remove harmful join rules from a Warded Datalog$^\pm$ set of rules and produce an equivalent Harmless Warded Datalog$^\pm$ set. First, we present the overall algorithm, then we discuss each main step in greater detail.

\noindent
\textbf{Harmful Join Elimination (HJE) Algorithm}.
Let $\rho$ be a generic harmful join rule in the following form:
\begin{center}
    $\rho:\forall x\forall y\forall{\overline h}(A(x_1,y_1,{\overline h}),B(x_2,y_2,{\overline h})\to\exists z\,C(x,z))$.
\end{center}
Let $\Sigma$ be a set of Warded Datalog$^\pm$ rules containing $\rho$ and $\Gamma$ be the generic multiset of causes of affectedness for $\rho$.\\[2mm]
\texttt{Cause Elimination}. Until there are harmful join rules $\rho$ in $\Sigma$:

\begin{enumerate}[noitemsep,nolistsep]
    \item \textit{Grounding}. Add to $\Sigma$ the following rules:\\
    Dom$({\overline h}),A(x_1,y_1,{\overline h}) \to A^\prime(x_1,y_1,{\overline h})$\\
    $A^\prime(x_1,y_1,{\overline h}) \to A(x_1,y_1,{\overline h})$\\
    $A^\prime(x_1,y_1,{\overline h}),B(x_2,y_2,{\overline h})\to\exists z\,C(x,z))$
    \item \textit{Recognition} (\texttt{Folding}). For each harmful join rule\\ $\beta: A^\prime(x_1,y_1,{\overline h}), B^\prime(x_2,y_2,{\overline h}) \to \exists z\,C^\prime(x,z))$ distinct from $\rho$ such that there exists a body-isomorphism with $\rho$, add to $\Sigma$ rule $C(x,z^\prime) \to \exists z\,C^\prime(x,z)$, remove $\beta$ and jump to step 4.
    \item \textit{Back-Composition}. Let \textit{T} be a new HU-Tree with $\rho$ as root. For each $\Gamma$ and until $\rho$ has causes of affectedness in current $\Gamma$, compose back along them and build \textit{T} as follows.
    \begin{enumerate}[noitemsep,nolistsep]
        \item \textit{Indirect Back-Composition} (\texttt{Unfolding}).\\
        For each indirect cause $D(x,y,{\overline h})\to\exists z\, A(x,z,{\overline h})$\\
        remove the cause from $\Gamma$ and add to \textit{T} the rule\\ 
        $D(x_1,y_1,{\overline h}),B(x_2,y_2,{\overline h}) \to \exists z\,C(x,z)$
        \item \textit{Ancestor} \texttt{Folding}. For each rule $\rho^\prime$ such that there exists (partial) body-isomorphism with one of its ancestors $\rho$, add to \textit{T} a new node labelled by the result of folding $\rho^\prime$ into $\rho$ and jump to step 4.
        \item \textit{Direct Back-Composition} (\texttt{Existential Unfolding}).\\
        For each direct cause $\forall x\forall y\,D(x,y) \to \exists z\exists h\,A(x,z,{\overline h})$\\
        remove the cause from $\Gamma$ and add to \textit{T} the rule \\ 
        $D(x_1,y_1), B(x_2,y_2,f_\beta(x_1,y_1)) \to \exists z\,C(x,z)$
    \end{enumerate}
    \item add the leaves of \textit{T} to $\Sigma$.
    \item mark \textit{$\rho$} as removed from $\Sigma$.
\end{enumerate}
\noindent
\texttt{Skolem-Simplification}. For each rule added to $\Sigma$ at the previous step, depending on its form, proceed as follows:
\begin{enumerate}[noitemsep,nolistsep]
\item \textit{Virtual Joins}: Remove from $\Sigma$ rules of the form:
\begin{enumerate}[noitemsep,nolistsep,label=\alph*.]
\item $A(x_1,y_1,h),B(x_2,y_2,f_\beta(h)),... \to \exists z\,C(x,z)$
\item $A(x_1,y_1,h=f_{\beta_1}(\cdot)),B(x_2,y_2,h=f_{\beta_2}(\cdot)),... \to \exists z\,C(x,z)$
\item $A(x_1,y_1,h=f_\beta(\cdot)),B(x_2,y_2,h=f_\beta(f_\beta(...\cdot))) \to \exists z\,C(x,z)$
\end{enumerate}
\item \textit{Linearization}: Unify into $\Sigma$ rules of the form:\\
$A(x_1,y_1,h=f_\beta(\cdot)),A(x_2,y_2,h=f_\beta(\cdot)),... \to \exists z\,C(x,z)$, into \textit{$A(x,y,h),... \to \exists z\,C(x,z)$}.
\end{enumerate}

\smallskip
\noindent
Starting from a pseudocode version of HJE (\textbf{Algorithm~\ref{alg:hje}}), let us now analyse each step in detail.

\begin{algorithm}
\footnotesize
\caption{\texttt{HarmfulJoinElimination(SetRules $\Sigma$)}}
\label{alg:hje}
\begin{algorithmic}[1]
\State{\textcolor{darkblue}{\{detach from danger the harmful join rules\}}}
\State {\Call {\textcolor{darkgreen}{Grounding}}{$\Sigma$}} \newline
\textcolor{darkblue}{\{attempt removing harmful join rules without composition\}}
\State {\Call {\textcolor{darkgreen}{Recognition}}{$\Sigma$}} \newline
\textcolor{darkblue}{\{compose back along the causes of affectedness\}}
\State {\Call {\textcolor{darkgreen}{Back-Composition}}{$\Sigma$}} \newline
\textcolor{darkblue}{\{manage rules which contain Skolem functions\}}
\State {\Call {\textcolor{darkgreen}{Skolem-Simplification}}{$\Sigma$}}
\end{algorithmic}
\end{algorithm}

\smallskip
\noindent
\textbf{Algorithm~\ref{alg:grounding}} (\textit{Grounding}) produces a ground harmless copy of the harmful join rule: this copy operates only on EDB facts thanks to the Dom feature, which compels the harmful variables of the atom \textit{a} to bind only to constants in the domain, thus avoiding the propagation of nulls. Moreover, Grounding considers all the other rules in $\Sigma$ that are not causes of affectedness for the harmful join rule and if their head unifies with the atom \textit{a}, it is accordingly renamed to \textit{a$^\prime$}. The causes are not modified and will be considered by the \textit{Back-Composition} phase (Algorithm~\ref{alg:backcomposition}).

\begin{algorithm}
\footnotesize
\caption{\texttt{Grounding(SetRules $\Sigma$)}}
\label{alg:grounding}
\begin{algorithmic}[1]
\State {rulesAndCauses $=$ \textcolor{darkgreen}{getHHJoinAffectednessCausesByRule}($\Sigma$)} \newline
\textcolor{darkblue}{\{let a(*) be an atom of R involved in the HH join\}}
\For {Rule $\mathcal{R}$ in rulesAndCauses} \newline
    \textcolor{darkblue}{$\enspace$ $\enspace$\{add Rule a$^\prime$(*) :- a(*), Dom(*) to $\Sigma$\}} \newline
    \textcolor{darkblue}{$\enspace$ $\enspace$\{add Rule a(*) :- a$^\prime$(*) to $\Sigma$\}}
    \State {\textcolor{darkgreen}{addGroundingRules}($\Sigma$)}\newline
    \textcolor{darkblue}{$\enspace$ $\enspace$\{replace $\mathcal{R}$ renaming a(*) with a$^\prime$(*)\}} \newline
    \textcolor{darkblue}{$\enspace$ $\enspace$\{rename head of rules non-causes whose head unifies\}}
    \State {\textcolor{darkgreen}{renameRulesNonCauses}($\Sigma$)}
\EndFor
\State {\textbf{end}}
\end{algorithmic}
\end{algorithm}

\smallskip
\noindent
\textbf{Algorithm~\ref{alg:recognition}} (\textit{Recognition}) attempts \texttt{folding} between each pair of harmful join rules. Its purpose is to avoid the application of the \textit{Back-Composition} phase (Algorithm~\ref{alg:backcomposition}) on rules which are body-isomorphic to others, thus limiting the number of HU-Trees built and so the steps required for the algorithm to terminate. As two body-isomorphic harmful join rules in the same set of rules have the same causes of affectedness, the new rules added by the \textit{Back-Composition} also have isomorphic bodies: the implication added by the \textit{Recognition} avoids any loss of meaning.

\begin{algorithm}
\footnotesize
\caption{\texttt{Recognition(SetRules $\Sigma$)}}
\label{alg:recognition}
\begin{algorithmic}[1]
\State {hhJoinRules = \textcolor{darkgreen}{getHHJoinRules}($\Sigma$)} \newline
\textcolor{darkblue}{\{check for body-isomorphism between harmful join Rules\}}
\For {($\mathcal{R}$1,$\mathcal{R}$2) in hhJoinRules} \newline
    \textcolor{darkblue}{$\enspace$ $\enspace$\{if $\mathcal{R}$1 and $\mathcal{R}$2 are body-isomorphic\}}
    \If {\textcolor{darkgreen}{areBodyIsomorphic}($\mathcal{R}$1,$\mathcal{R}$2)} \newline
        \textcolor{darkblue}{$\enspace$ $\enspace$\ $\enspace$\ $\enspace$\{remove $\mathcal{R}$2\}} \newline
        \textcolor{darkblue}{$\enspace$ $\enspace$\ $\enspace$\ $\enspace$\{add rule built as follows: $\mathcal{R}$2.head :- $\mathcal{R}$1.head\}}
        \State {\textcolor{darkgreen}{folding}($\mathcal{R}$1,$\mathcal{R}$2)}
    \EndIf
    \State {\textbf{end}}
\EndFor
\State {\textbf{end}}
\end{algorithmic}
\end{algorithm}

\smallskip
\noindent
\textbf{Algorithm~\ref{alg:backcomposition}} (\textit{Back-Composition})
is the central phase of HJE. Given a harmful join rule, it individuates, for each harmful variable of an atom in the harmful join, all the \textit{causes of affectedness}; a new HU-Tree is also built with the harmful join rule as root. The following process is performed iteratively as long as $\Sigma$ contains harmful joins and, for each harmful join rule, until the Back-Composition reaches the direct cause of affectedness. First, it \texttt{unfolds} the current rule with its causes, in the \texttt{existential} variant in case of direct causes; then, it checks whether there is a full or partial body-isomorphism with one of its ancestors in the current branch of the HU-Tree and it attempts \texttt{folding}. Finally, it updates the corresponding HU-Tree and, if the folding succeeded, the procedure terminates and its corresponding branch is cut (as shown in Section~\ref{hjeTheoreticalResult}). When all the harmful join rules have been back-composed with their causes of affectedness, up to the direct ones, they are removed and the resulting rules (that is, the leaves in the HU-Trees) added to $\Sigma$.

\begin{algorithm}
\footnotesize
\caption{\texttt{Back-Composition(SetRules $\Sigma$)}}
\label{alg:backcomposition}
\begin{algorithmic}[1]
\State {rulesAndCauses = \textcolor{darkgreen}{getAffectednessCausesByRule}($\Sigma$)} \newline
\textcolor{darkblue}{\{define new HU-Tree for each harmful join rule HR\}}
\State {huTree = new HU-Tree(HR)} \newline
 \textcolor{darkblue}{\{for each rule R that contains HH join or\newline affected variable constrained by Skolem atom\}}
\While {!rulesAndCauses.\textcolor{darkgreen}{isEmpty}()}
    \State {Rule $\mathcal{R}$ = rulesAndCauses.\textcolor{darkgreen}{nextRule}()} \newline
    \textcolor{darkblue}{$\enspace$ $\enspace$\{for each cause of affectedness\}}
    \For {Rule cause in $\mathcal{R}$.\textcolor{darkgreen}{getCauses}()} \newline
        \textcolor{darkblue}{$\enspace$ $\enspace$ $\enspace$ $\enspace$ \{try unfolding, handling existentials\}}
        \State {Rule $\mathcal{RN}$ = \textcolor{darkgreen}{existentialUnfolding}($\mathcal{R}$,cause)} \newline
        \textcolor{darkblue}{$\enspace$ $\enspace$ $\enspace$ $\enspace$\{try folding with ancestors of $\mathcal{R}$\}}
        \State {\textcolor{darkgreen}{ancestorFolding(}$\mathcal{RN}$,huTree)} \newline
        \textcolor{darkblue}{$\enspace$ $\enspace$ $\enspace$ $\enspace$\{update the HU-Tree\}}
        \State {huTree.\textcolor{darkgreen}{update}($\mathcal{RN}$,$\mathcal{R}$)}
    \EndFor
    \State {\textbf{end}} \newline
    \textcolor{darkblue}{$\enspace$ $\enspace$\{update rulesAndCauses\}}
\EndWhile
\State {\textbf{end}} \newline
\textcolor{darkblue}{\{update $\Sigma$ adding HU-Tree leaves and removing $\mathcal{R}$\}}
\State {$\Sigma$.\textcolor{darkgreen}{update}($\mathcal{R}$,hutree)}
\State {\textbf{end}}
\end{algorithmic}
\end{algorithm}

\smallskip
\noindent
\textbf{Algorithm~\ref{alg:skolemsimplification}} (\textit{Skolem-Simplification}) handles the rules that contain Skolem atoms after \textit{Back-Composition} (Algorithm~\ref{alg:backcomposition}) and produces the final set of rules returned by HJE. It considers two cases.

In the \textit{Virtual Joins} case, such rules are dropped as the conditions on the Skolem functions cannot be satisfied for one of the following reasons: (i)~a harmless variable is equated to a Skolem function, unsatisfiable by definition;
(ii)~two distinct Skolem functions are equated, unsatisfiable by range disjointness of Skolem functions; (iii)~a Skolem function is equated to its recursive application, which is also unsatisfiable by injectivity of Skolem functions.

In the \textit{Linearization} case, a rule involving Skolem functions is simplified in such a way that two atoms with the same Skolem function are unified, and the function itself is replaced by a variable.

\begin{algorithm}
\footnotesize
\caption{\texttt{Skolem-Simplification(SetRules $\Sigma$)}}
\label{alg:skolemsimplification}
\begin{algorithmic}[1]
\State {bucket = \textcolor{darkgreen}{getRulesFromBackComposition}($\Sigma$)}
\For {Rule $\mathcal{R}$ in bucket} \newline
    \textcolor{darkblue}{$\enspace$ $\enspace$\{remove virtual join cases:\}} \newline
    \textcolor{darkblue}{$\enspace$ $\enspace$\{- Skolem atom conditions unaffected variable in body\}} \newline
    \textcolor{darkblue}{$\enspace$ $\enspace$\{- single Skolem atom conditions variable not in body\}} \newline
    \textcolor{darkblue}{$\enspace$ $\enspace$\{- recursive Skolem atom\}}
    \If {\textcolor{darkgreen}{isVirtualJoinCase}($\mathcal{R}$)}
        \State {$\Sigma$.\textcolor{darkgreen}{dropRule}($\mathcal{R}$)} \newline
    \textcolor{darkblue}{$\enspace$ $\enspace$\{linearization case:\}} \newline
    \textcolor{darkblue}{$\enspace$ $\enspace$\{- multiple Skolem atoms condition variable not in body\}}
    \Else
        \If {\textcolor{darkgreen}{isLinearizationCase}($\mathcal{R}$)} \newline
            \textcolor{darkblue}{$\enspace$ $\enspace$\ $\enspace$\{if unify: drop and propagate unification to whole rule\}} \newline
            \textcolor{darkblue}{$\enspace$ $\enspace$\ $\enspace$\{if do not unify: drop rule\}}
            \State {\textcolor{darkgreen}{attemptUnification}($\mathcal{R}$)}
        \EndIf
        \State {\textbf{end}}
    \EndIf
    \State {\textbf{end}}
\EndFor
\State {\textbf{end}} \newline
\textcolor{darkblue}{\{deduplicate rules\}}
\State {$\Sigma$.\textcolor{darkgreen}{deduplicateRules}()}
\end{algorithmic}
\end{algorithm}
\label{hjeAlgorithm}
\subsection{Algorithm Correctness and Properties}
\noindent
We conclude the section by discussing the correctness of our HJE as well as other relevant theoretical aspects.

\begin{theorem}
\label{th:hjecorrectness}
Let $\Sigma$ be a Warded Datalog$^\pm$ set of rules and $\rho_1,...,\rho_n$ be harmful join rules in $\Sigma$. Then \textnormal{HJE}($\Sigma$) produces an equivalent (i.e., meaning-preserving) Harmless Warded Datalog$^\pm$ set of rules.
\end{theorem}

\begin{proof}
The HJE algorithm generates the HU-Trees we have seen in the proof of Theorem~\ref{th:hjeharmlesswarded}. Specifically, \textit{Back-Composition} applies folding and unfolding to the harmful join rules up to the direct causes of affectedness, thus producing, by construction, (up to) \textit{n} HU-Trees. \textit{Recognition} applies folding, preventing the creation of isomorphic trees. \textit{Grounding} adds three rules to produce a ground harmless copy of $\rho_i$, as also shown in the aforementioned proof. Finally, \textit{Skolem-Simplification} simplifies Skolem atoms previously added and attempts unification between the leaves of the trees, dropping repeated rules and rules which would never activate: thus, by definition, no transformation breaks the equivalence of the normalized set of rules. Therefore, we can conclude that the HJE algorithm produces an equivalent Harmless Warded Datalog$^\pm$.
\end{proof}

\noindent
\textbf{Role of Folding}. Folding is essential to avoid non-termination of the algorithm in two cases: direct or indirect recursive rules and multiple mutually recursive rules as causes of affectedness. Figure~\ref{fig:hjeexecution} shows the partial HU-Tree corresponding to the Back-Composition phase on a set of recursive  Vadalog rules, inspired by an evaluation scenario for the Vadalog system~\cite{bellomarini2018vadalog}. Specifically, Figure~\ref{fig:hjeexecution}(a) illustrates the first steps of the execution, which does not terminate because of the recursion on the indirect cause of affectedness $\beta$. Figure~\ref{fig:hjeexecution}(b) shows how folding prevents non-termination by pruning the tree when it detects body-isomorphism between the current rule and one of its ancestors. The red letters denote variables in affected positions. The resulting full set of rules is in the Appendix~\ref{appendixHJE}.

\smallskip
\noindent
\textbf{Complexity}. Based on Proposition~\ref{prop:hjetreecomplexity}, we have that 
the HJE algorithm runs in time exponential in \texttt{mdh}, as \textit{Back-Composition} takes at most as many steps as the ones required by the procedure used in the proof of Theorem~\ref{th:hjeharmlesswarded} to build the HU-Trees. Figure~\ref{fig:hjerelevantcases} shows relevant examples. 
Specifically, Figure~\ref{fig:hjerelevantcases}(a) shows the application of HJE with $k$=$d$=2 and the final rules in the harmless set after \textit{Skolem-Simplification};
Figure~\ref{fig:hjerelevantcases}(b) shows the effect of recursion within the sequence of causes of affectedness. From the example, it is clear that the number of folding checks ---only the successful ones are reported--- is exponential in \texttt{mdh}.
\label{hjeCorrectness}
\section{Experimental Evaluation}
\noindent
In this section we illustrate the experimental evaluations, organized as follows.
In Section~\ref{testingiWarded}, we study the performance of {\sc iWarded} in dependence on the most relevant input parameters. In Section~\ref{testingiBench} we briefly recall the metadata generator {\sc iBench} and provide a comparison analysis with {\sc iWarded}. In Section~\ref{testingHJE} we show our algorithm for HJE in action and study how the characteristics of the input rules, in Harmful Warded Datalog$^\pm$, affect the runtime. We close the circle by evaluating HJE runtime on synthetic scenarios purposely generated with {\sc iWarded}.

\smallskip
\noindent
\textbf{Technical Notes}. Both {\sc iWarded} and HJE are implemented in Java~13. Experiments have been run on an Asus Zenbook UX331UN with 4 x 2.78 GHz CPU and 8GB of RAM.

\begin{figure*}[!bth]
  \centering
  \includegraphics[width=.90\textwidth]{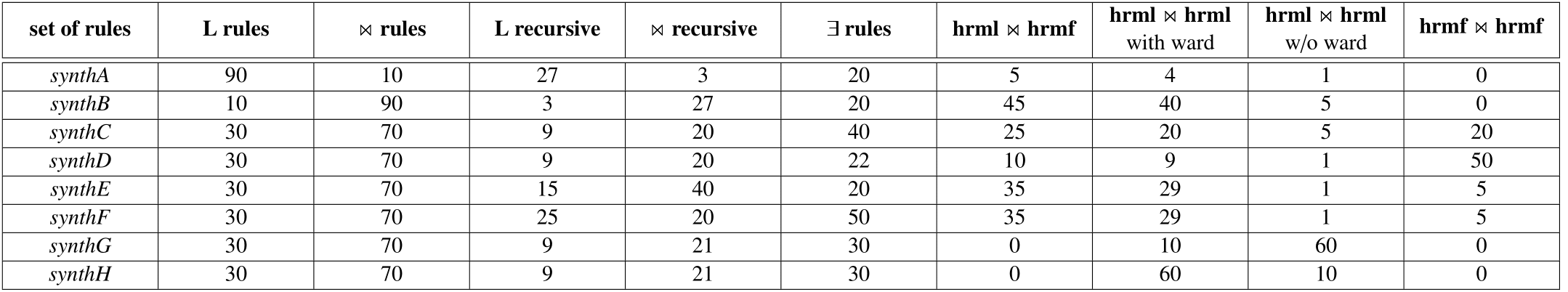}
  \caption{Details of the scenarios generated with  \textbf{\textsc{iWarded}}.}
 \label{fig:synth}
\end{figure*}
\begin{figure*}[!htb]
\centering
 \begin{tabular}{@{}ccc@{}}
    \subfloat[{ \textsc{iWarded}: Structural scenarios.}]{%
    \includegraphics[width=.33\textwidth]{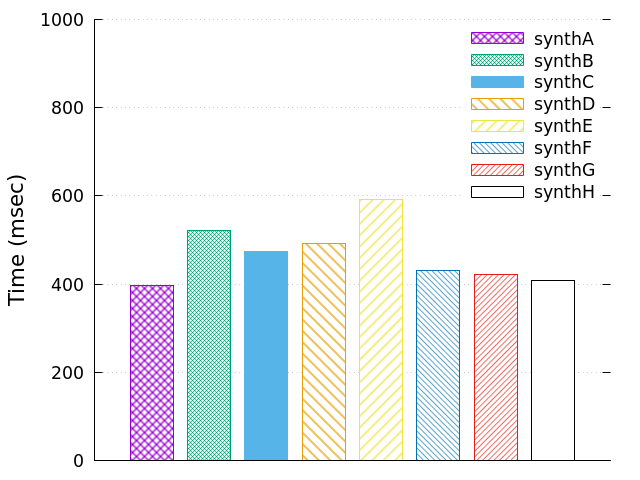}}\hfill
    \subfloat[{ \textsc{iWarded}: Scalability scenario.}]{%
    \includegraphics[width=.33\textwidth]{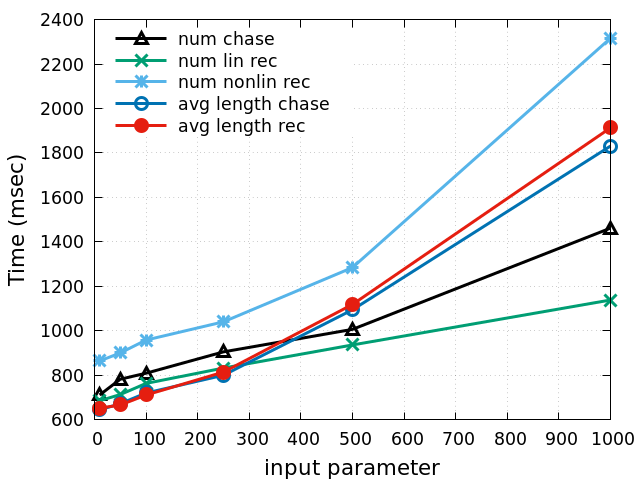}}
    \subfloat[{ Generation times \textsc{iWarded}-\textsc{iBench}.}]{%
    \includegraphics[width=.33\textwidth]{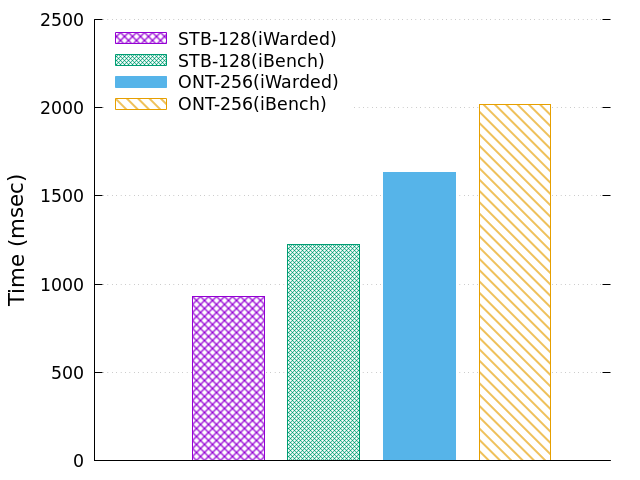}}
 \end{tabular}
 \begin{tabular}{@{}ccc@{}}
    \subfloat[{HJE: Runtime varying the \# of harmful joins.}]{%
    \includegraphics[width=.33\textwidth]{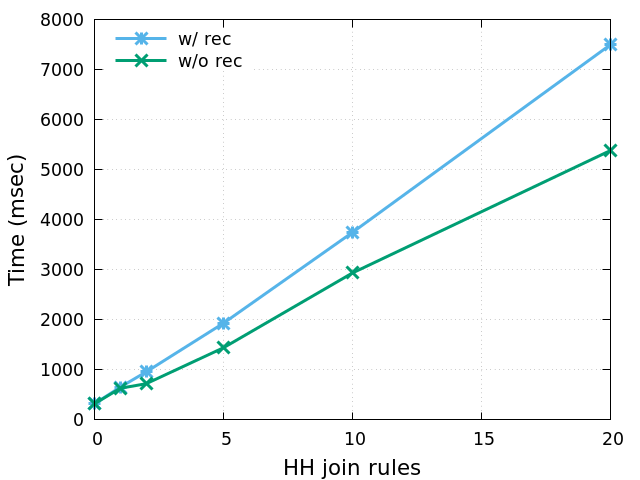}}
    \subfloat[{HJE: Runtime with \textsc{iWarded} scenarios.}]{%
    \includegraphics[width=.33\textwidth]{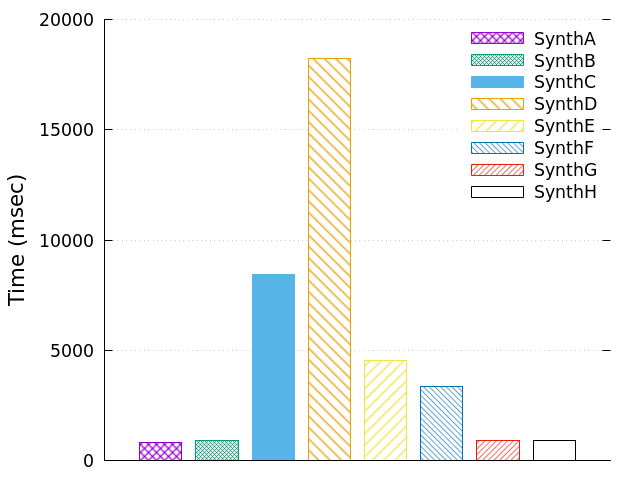}}
    \subfloat[{HJE: \# of rules varying HU-Tree depth ($d$) and branching factor ($k$).}]{%
    \includegraphics[width=.33\textwidth]{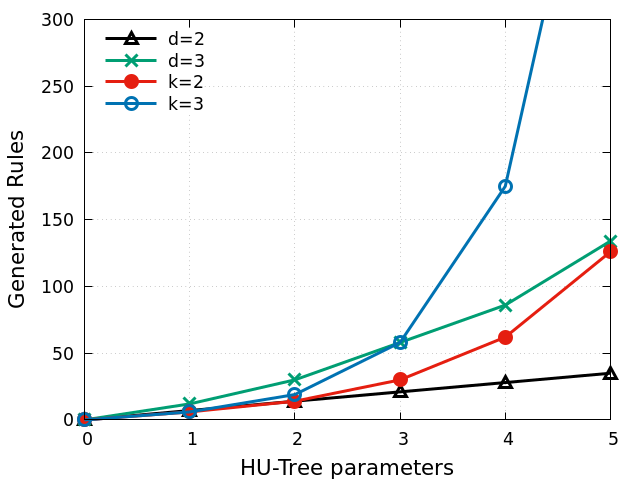}}
 \end{tabular}
 \caption{Results for the experimental scenarios.}
 \label{fig:testing}\smallskip\smallskip\smallskip
\end{figure*}
\label{testing}
\subsection{iWarded Evaluation}
\noindent
We verified the performance of {\sc iWarded} in generating sets of rules (we neglect the time needed to generate synthetic input data as uninteresting in this setting) in several complex scenarios, with the goal of understanding how the different structures, induced by the parameters in Figure~\ref{fig:cases}, affect the generation process.
Note that in this evaluation we do not focus our attention on the performance of the reasoning tasks in the different settings, for which the reader is referred to the Appendix and to the full evaluation of the Vadalog system~\cite{bellomarini2018vadalog}.

\smallskip
The first set of tests (Figure~\ref{fig:testing}(a)) consists of a group of eight  \textit {structural scenarios}, whereas the second set (Figure~\ref{fig:testing}(b)) refers to a single \textit {scalability scenario} with respect to relevant input parameters: we analyse the results in terms of execution time for the generation. For each scenario we ran 50 tests and averaged the resulting times. 

\smallskip
\noindent
\textbf{Structural Scenarios}. Each scenario consists of 100 rules, distributed among linear, existential and join as follows:
\begin{itemize}[noitemsep,nolistsep,leftmargin=2mm]
	\item \textit{synthA} shows a prevalence of linear rules, 20\% having existential quantification; 30\% of linear and non-linear rules are recursive and joins are equally distributed among harmless-harmful and harmless-harmless, with a prevalence of joins with ward.
    \item \textit{synthB} is a variant of \textit{synthA} with more non-linear rules.
    \end{itemize}
\begin{itemize}[noitemsep,nolistsep,leftmargin=2mm]
	\item \textit{synthC} as well as all the following scenarios have 30\% of linear rules and 70\% of non-linear rules.
	\item \textit{synthD} is a variant of \textit{synthC} with more harmful-harmful joins.
\end{itemize}
\begin{itemize}[noitemsep,nolistsep,leftmargin=2mm]
	\item \textit{synthE} and \textit{synthF} aim to verify the impact on the performance of a relevant presence of recursion, the first on join rules, the second on linear rules.
	\item \textit{synthG} and \textit{synthH} are similar to \textit{synthC} and \textit{synthD}, but have more harmless-harmless joins, w/o and with ward, respectively.
\end{itemize}
Figure~\ref{fig:synth} fully reports the associated input parameters.

\smallskip
\noindent
\textbf{Scalability Scenario}. We generated sets of rules by increasing (10, 50, 100, 250, 500 and 1000) the value of each of the following parameters, while keeping the others unvaried. In particular, we acted on: (1)~the number of input-output sequences; (2)~the number of linear recursive sequences; (3)~the number of non-linear recursive sequences; (4)~the average length for input-output sequences; (5)~the average recursion length.

\smallskip
\noindent
\textbf{Results}. 
The execution times, reported in Figure~\ref{fig:testing}(a), for the \textit{structural scenarios} confirm the effectiveness of the sequence network abstraction adopted in {\sc iWarded}. As there is no back-propagation of affected positions, rules need not be updated and so elapsed time is simply proportional to the number of intended chase steps, as specified as an input parameter (see Figure~\ref{fig:cases}). In absolute terms, performance is very satisfactory, always under 1~sec. 
Moreover, {\sc iWarded} runtime is robust to structural variations. 
The longest times are exhibited by \textit{synthA} and \textit{synthE}. This is directly motivated by the higher rate of recursive sequences with non-linear joins requiring a higher number of extra rules to close the recursion (one for each body atom). Nevertheless, also this case confirms the dependence of runtime on the number of rules to be generated.
Figure~\ref{fig:testing}(b) shows the average times required for the generation of a set of rules when varying the input parameters. The trends confirm polynomial (slightly superlinear) behaviour of the algorithm. Indeed, the average length of recursive sequences has greater impact on the elapsed times, as observed for \textit{SynthB} and \textit{SynthE}.
\label{testingiWarded}
\subsection{iBench and iWarded}
\noindent
{\sc iBench} is a popular benchmark generator for data integration and data exchange settings specified in terms of \textit{tuple-generating dependencies} (TGDs)~\cite{arocena2015ibench}. {\sc iBench} and {\sc iWarded} have different application context and target: {\sc iBench} has the goal to generate very large and realistic settings, does not put emphasis on the evaluation strategies adopted by the target systems and is based on the simple language of TGDs, without recursion; {\sc iWarded} has full focus on Datalog$^\pm$ and emphasizes the possibility to control many aspects and features of the rules, with attention to recursion, so as to evaluate execution strategies in target systems such as reasoners.

\smallskip
\noindent
\textbf{Test Scenarios}. In this section we considered the specifications for two test scenarios, \textit{STB-128} and \textit{ONT-256}, fully described by Benedikt et al.~\cite{benedikt2017benchmarking}, and instructed both {\sc iWarded} and {\sc iBench} to generate rules and TGDs for them, respectively. For each scenario we performed 50 runs and compared the average execution time.

\smallskip
\begin{itemize}[noitemsep,nolistsep,leftmargin=2mm]
\item \textit{STB-128}: a set of about 250 warded rules, 25\% of which contain existentials, with 15 harmful joins, 30 cases of propagation of labelled nulls with warded rules and 112 distinct predicates;
\item \textit{ONT-256}: a set of 789 warded rules, 35\% of which contain existentials, with 295 harmful joins, more than 300 cases of propagation of labelled nulls with warded rules and 220 distinct predicates.
\end{itemize}

\smallskip
\noindent
\textbf{Results}. Figure~\ref{fig:testing}(c) shows that {\sc iWarded} performs better than {\sc iBench} for both STB-128 and ONT-256, taking 900 vs 1200 milliseconds and 1600 vs 2000 ms, respectively. This result confirms that in standard but realistic settings, combining rule templates with a memory-free approach (see Section~\ref{iWardedCorrectness}) is time effective.
\label{testingiBench}
\subsection{Evaluation of Harmful Join Elimination}
\noindent
We evaluated the HJE algorithm presented in Section~\ref{hjeAlgorithm} in the following three scenarios. For the first two scenarios we performed 50 runs each, measuring the average generation time, and for the third one we performed 20 executions. 

\smallskip
\noindent
\textbf{HH Join Rules and Recursion}. We built multiple versions of a single set of 20 rules, by gradually increasing the number of rules having harmful joins, with and without recursive causes. We made sure to avoid body-isomorphic rules to prevent simplifications by Recognition. We generated instances with 1, 2, 5, 10 and 20 harmful join rules (so from 21 to 40 rules in total) and measured the average execution time. Figure~\ref{fig:testing}(d) shows the linear behaviour of the HJE algorithm in the number of harmful join rules: this is a consequence of the procedure discussed in Section~\ref{hje}, since each harmful join rule causes the creation of a single HU-Tree
because of Recognition. The presence of recursion in the sequences of causes of affectedness impacts performance, due to the additional folding checks in the execution, e.g., with $7.5$ vs.\ $5.2$ seconds for 20 harmful join rules.

\smallskip
\noindent
\textbf{HJE on \textsc{iWarded} Scenarios}. We ran the HJE algorithm on all the scenarios generated by {\sc iWarded} that we have seen in Section~\ref{testingiWarded} and measured the average time required for normalization. As reported in Figure~\ref{fig:synth}, the settings containing harmful joins are \textit{synthE} and \textit{synthF} (5), \textit{synthC} (20), and \textit{synthD} (50). The results are reported in Figure~\ref{fig:testing}(e). In the absence of harmful joins, the execution time reflects the time needed to check that the input scenario is within the Harmless Warded Datalog$^\pm$ fragment. On the other hand, both \textit{synthE} and \textit{synthF} require normalization, which however takes less than 5 seconds, as their structure does not present long sequences of causes of affectedness. Interestingly, for \textit{synthC} and \textit{synthD} we observe 8 and 18 seconds, witnessing that more complex sets of rules (many existentials, and therefore many causes of affectedness), produce higher normalization times.

\smallskip
\noindent
\textbf{HJE with different HU-Trees}. We tested in practice the scenario originally presented in Section~\ref{hjeTheoreticalResult} and shown in Figure~\ref{fig:hjerelevantcases}(a). The results are reported in Figure~\ref{fig:testing}(f): we applied HJE to a set of rules with a single harmful join rule, having a HU-Tree of increasing depth $d$ and branching factor $k$ (i.e., the number of unfolding for each node of the tree). We varied each of the parameters from 0 to 5 while keeping the other constant (2 or~3). For each setting, we counted the number of generated rules (nodes in the tree) throughout the HJE execution. The experiment reflects the theoretical results of Proposition~\ref{prop:hjetreecomplexity}. In particular, we observe that the number of rules grows polynomially with $k$ and exponentially with $d$, as apparent for the curve with $k=3$. In conclusion, the runtime of HJE is directly dependent on the number of generated rules, which in turn is determined by the maximum distance from harmlessness (\texttt{mdh}), i.e., the ``effort'' in number of unfoldings needed to resolve all the direct and indirect causes of affectedness.
\label{testingHJE}
\section{Related Work}
\noindent
To the best of our knowledge, this is the first attempt to provide the logic-based reasoning community with a tool to generate tailored Datalog$^\pm$ benchmarks. 
Great efforts in empiric evaluation have been spent by both the database systems community (e.g., with TPC~\cite{tpcwebpage}) and the theorem proving one (e.g., with SMTLib~\cite{smtwebpage} and TPTP~\cite{tptpwebpage}). On the other hand, in the literature about extending data management tools with reasoning features we do not find equivalent richness, although evaluation of many systems has been carried out~\cite{BoIL16,GMPS14,KoAm14}. Our two main pillars are {\sc iBench}~\cite{arocena2015ibench}, the popular metadata generator for data integration settings we have discussed, and {\sc ChaseBench}~\cite{benedikt2017benchmarking}, a recent proposal for a comprehensive benchmarking suite for {\sc chase}-based systems. The main idea of {\sc iBench} is providing a fully controllable metadata generator, giving the user maximum customization capabilities. Yet there is no coverage for reasoning settings and only standard TGD-based schema mappings are supported. On the other hand, {\sc ChaseBench} offers an extremely valuable set of test scenarios, partially repurposing those from {\sc iBench}, and its main contribution lies in making refined test scenarios available and in taking the burden to carefully test the most relevant reasoners. Yet, test generation capabilities and Datalog$^\pm$ are not its focus: {\sc ChaseBench} can generate weakly acyclic TGDs, the standard decidable setting for the {\sc chase}, allowing to control the chase depth and overall TGD complexity. With {\sc iWarded} we want to be complementary, possibly even contributing to enriching the {\sc ChaseBench} toolbox. The goal of our work has been to fully develop the user control idea of {\sc iBench} into the specificities of the Datalog$^\pm$ fragments. In these settings, termination conditions go beyond weak acyclicity and require non-trivial normalization, and countless aspects deriving from the interplay between existential quantification and recursion must be taken into account and made tunable by the user.

Existing benchmarks such as LUBM~\cite{guo2005lubm}, STB~\cite{AlCV08} and ONT~\cite{arocena2015ibench,benedikt2017benchmarking} pay special care to query scenarios, and {\sc ChaseBench} even provides a query generation tool. {\sc iWarded} follows a different strategy: as the rule characteristics can be fully controlled, it defines what rules are considered as output, and simulates conjunctive queries as rule bodies, according to the definition of reasoning task in Section~\ref{vadalog}.

We just touched on the broad topic of data generation, which is beyond the scope of this work. While some systems like {\sc ToXgene}~\cite{BMKL02} concentrate on this aspect of benchmarking, and currently {\sc iWarded} simply produces instances that stimulate specific query processing aspects (e.g., selectivity), instance generation in the reasoning realm requires more complex and often domain-dependent considerations, for example when specific graph topologies need to be simulated~\cite{HiBa08}, as we have experienced in applying reasoning in financial context~\cite{ABIS20}.

Many reasoning systems can be considered {\sc iWarded} target. Besides the Vadalog~\cite{bellomarini2018vadalog} system, specifically motivating the research, systems able to operate data integration and data exchange settings can certainly benefit from a large subset of the scenarios that can be generated by {\sc iWarded}: {\sc LLunatic}~\cite{geerts2014s,GMPS14}, {\sc Graal}~\cite{baget2015graal}, {\sc RDFox}~\cite{motik2014parallel}, {\sc PDQ}~\cite{BeLT14,BeLT142}, and  {\sc DLV}~\cite{LPFE06} are good examples. We also hope that the {\sc iWarded} benchmark will make it easier for more systems to support the full extent of Warded Datalog$^\pm$.
\label{relatedWork}
\section{Conclusion}
\noindent
With our work we took a step forward in providing the logic-based reasoning community with workable tools to generate tailored benchmarks for reasoning systems.
The recent introduction of a chase-dedicated benchmark in the database community has spurred us towards going beyond and leveraging the experience made in the data integration and data exchange literature with {\sc iBench}. In this wake, {\sc iWarded} addresses a relevant logic fragment, with interesting theoretical properties. In fact, the decidability of Warded Datalog$^\pm$ can be exploited within efficient and correct query answering algorithms when the rules are translated into a normalized, harmless form, that forbids joins on labelled nulls.  Our contribution includes both the theoretical results and the normalization algorithm, an essential element for {\sc iWarded}.

We plan to distribute the tool in the automated reasoning community and to keep developing it, with a focus on new features,
such as the generation of equality-generating dependencies (EGDs), and other existing or uprising logic fragments (such as Guarded, where all the affected variables are comprised within a single guard atom, or Piecewise Linear Datalog$^\pm$, having a more limited form of recursion yet supporting high parallelism) in the hope our work will simplify the empiric evaluation of many reasoning systems.

\noindent
\textbf{Additional notes}: The current release of the {\sc iWarded} generator can be found here~\cite{iwarded}.
\label{conclusion}

\balance

\bibliographystyle{ACM-Reference-Format}
\bibliography{biblio}

\appendix
\begin{appendices}
\section{Generator \lowercase{i}Warded}
\noindent
With reference to Section~\ref{iWarded} in the paper, we here provide additional details regarding {\sc iWarded} and the generation algorithm at its basis.

\smallskip
\noindent
\textbf{Annotation and Generated Set of Rules}. Among the Vadalog extensions supported by {\sc iWarded}, we also mention \textit{Annotations}. They are special facts that allow injecting specific behaviors into the Vadalog set of rules; specifically, {\sc iWarded} adopts the following. \textit{Input} annotations, which specify that the facts for an atom are imported from an external data source, such as a relational database: these are called \textit{extensional database} atoms (\textit{edb}), whereas the atoms which are not annotated are \textit{intensional database} atoms (\textit{idb}). \textit{Output} annotations, which specify that the facts for an atom are exported to an external destination, for example a relational database: these are called \textit{output intensional database} atoms (\textit{out}). \textit{Bind} annotations, which bind input or output atoms to a source.\\
Based on the concepts introduced above, in Figure~\ref{fig:iwardedprogramfull} we provide the set of rules generated with {\sc iWarded} and discussed in Section~\ref{iWardedAlgorithm}, here also showing the annotations for input and output atoms and the binding of the former to corresponding CSV files created during the generation process and based on the \textit{average selectivity} and the \textit{number of records in CSV files} parameters.

\setcounter{figure}{7}
\begin{figure}[hbt!]
    \centering
    \includegraphics[width=0.45\textwidth]{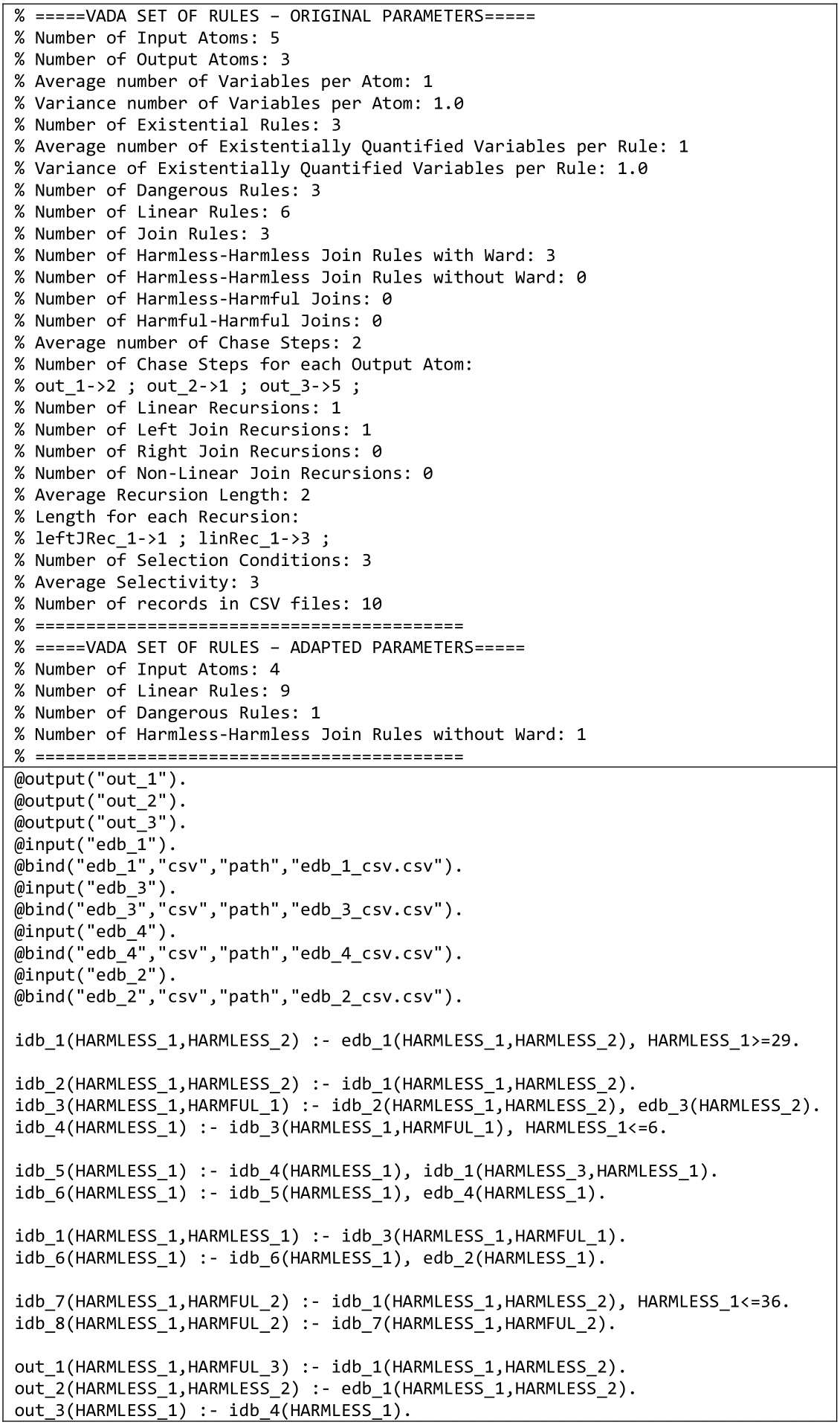}
    \caption{Set of Rules and Annotations with \textbf{\textsc{iWarded}.}}
    \label{fig:iwardedprogramfull}
\end{figure}

\smallskip
\noindent
\textbf{Generation Algorithm}. We now illustrate the algorithm for the generation of rules in the input-output sequence with {\sc iWarded}, with the goal of providing a deeper insight into the generation process. We then discuss the complexity of the iWardedGen algorithm.\\
\textbf{Algorithm~\ref{alg:genRuleInputOutput}} (\textit{GenerateRulesInputOutputSequence}) handles the creation of the \textit{input-output sequence}. Each rule is linked to the previous one, from the root up to when all the lengths of the sequences have been reached. Firstly, it selects whether the current rule is linear or join (and in case which type) and it builds the corresponding body, linked to the previous rule: this choice is random among the options which are allowed by the current status of the input scenario: for example, if the input value for the number of linear rules in the set has already been reached, then the rule will be a join one. Then it checks whether the body presents affected positions, propagated from the previous rules, and it selects the state of the rule to respect input requirements and preserve wardedness: for example, a harmful-harmful join rule shall not have the join variable in the head, as the rule would not be warded. Finally, it builds the actual rule, adding conditions randomly if the input parameters allow it, and it updates both chase and recursive requirements. This procedure is repeated until the longest sequence is built.

\setcounter{algorithm}{6}
\begin{algorithm}
\footnotesize
\caption{\texttt{GenerateRulesInputOutputSequence($\Sigma$,S)}}
\label{alg:genRuleInputOutput}
\begin{algorithmic}[1]
\While {!$\mathcal{S}$.inputOutputSequences.\textcolor{darkgreen}{isEmpty}()} \newline
	\textcolor{darkblue}{$\enspace$ $\enspace$\{type of rule, random but with respect to S\}}
    \State {type = \textcolor{darkgreen}{linearOrJoin}($\mathcal{S}$)} \newline
	\textcolor{darkblue}{$\enspace$ $\enspace$\{build body of current rule\}}
    \State {body = \textcolor{darkgreen}{generateRuleBody}(type,$\Sigma$)} \newline
    \textcolor{darkblue}{$\enspace$ $\enspace$\{check rule affectedness, based on link to previous ones\}}
    \State {isAffected = \textcolor{darkgreen}{checkAffectedness}(body,$\Sigma$)} \newline
    \textcolor{darkblue}{$\enspace$ $\enspace$\{choose rule state (dangerous...) respecting wardedness\}}
    \State {state = \textcolor{darkgreen}{selectRuleState}(body,isAffected,$\mathcal{S}$)} \newline
    \textcolor{darkblue}{$\enspace$ $\enspace$\{build head of current rule\}}
    \State {head = \textcolor{darkgreen}{generateRuleHead}(body,state)} \newline
    \textcolor{darkblue}{$\enspace$ $\enspace$\{build possible conditions of current rule\}}
    \State {conditions = \textcolor{darkgreen}{generateRuleConditions}(body,$\mathcal{S}$)} \newline
    \textcolor{darkblue}{$\enspace$ $\enspace$\{generate current rule\}}
	\State {$\mathcal{R}$ = new Rule(body,head,conditions)} \newline
	\textcolor{darkblue}{$\enspace$ $\enspace$\{add rule to set\}}
	\State {$\Sigma$.\textcolor{darkgreen}{addRule}($\mathcal{R}$)} \newline
	\textcolor{darkblue}{$\enspace$ $\enspace$\{update chase and recursive requirements\}} \newline
	\textcolor{darkblue}{$\enspace$ $\enspace$\{if all input-output sequences are complete\}}
    \For {seq in $\mathcal{S}$.inputOutputSequences}
	    \If {seq == 0}
	        \State {$\mathcal{S}$.inputOutputSequences.\textcolor{darkgreen}{remove}(seq)}
	    \EndIf
	\EndFor
	\State {\textbf{end}}
\EndWhile
\State {\textbf{end}}
\end{algorithmic}
\end{algorithm}

\setcounter{theorem}{2}
\begin{theorem}
\label{th:iwardgencomplexity}
The iWardedGen algorithm runs in time polynomial in $n$, where $n$ is the maximum length of a sequence in the set of rules based on input requirements.
\end{theorem}

\newpage

\begin{proof}
From the input scenario, reaching the compatibility of the parameters can be achieved in constant time. Generating the actual attributes for the sequences of rules and instantiating the corresponding primitives is linear in the number of the components (atoms, rules, recursions,...) to be created, which depends on the length of the sequences themselves. The remaining steps for rule generation follow a similar approach, thus it is here sufficient to discuss Algorithm~\ref{alg:genRuleInputOutput}. With reference to a single rule, the type selection, the rule creation and the scenario update require constant time, as they only depend on the previous rule and the current scenario. On the other hand, the check for affectedness is achieved in linear time in the number of rules already in the sequence: the memory-free generation approach adopted in {\sc iWarded} makes the generic affected position only dependent on the rules previously added and allows to avoid more complex memory structures. As discussed above, this procedure is repeated until the longest sequence (of length $n$, based on input requirements) is built.
\end{proof}

\label{appendixiWarded}
\section{Harmful Join Elimination}
\begin{figure*}[!hbt]
  \begin{center}
    \includegraphics[width=\textwidth]{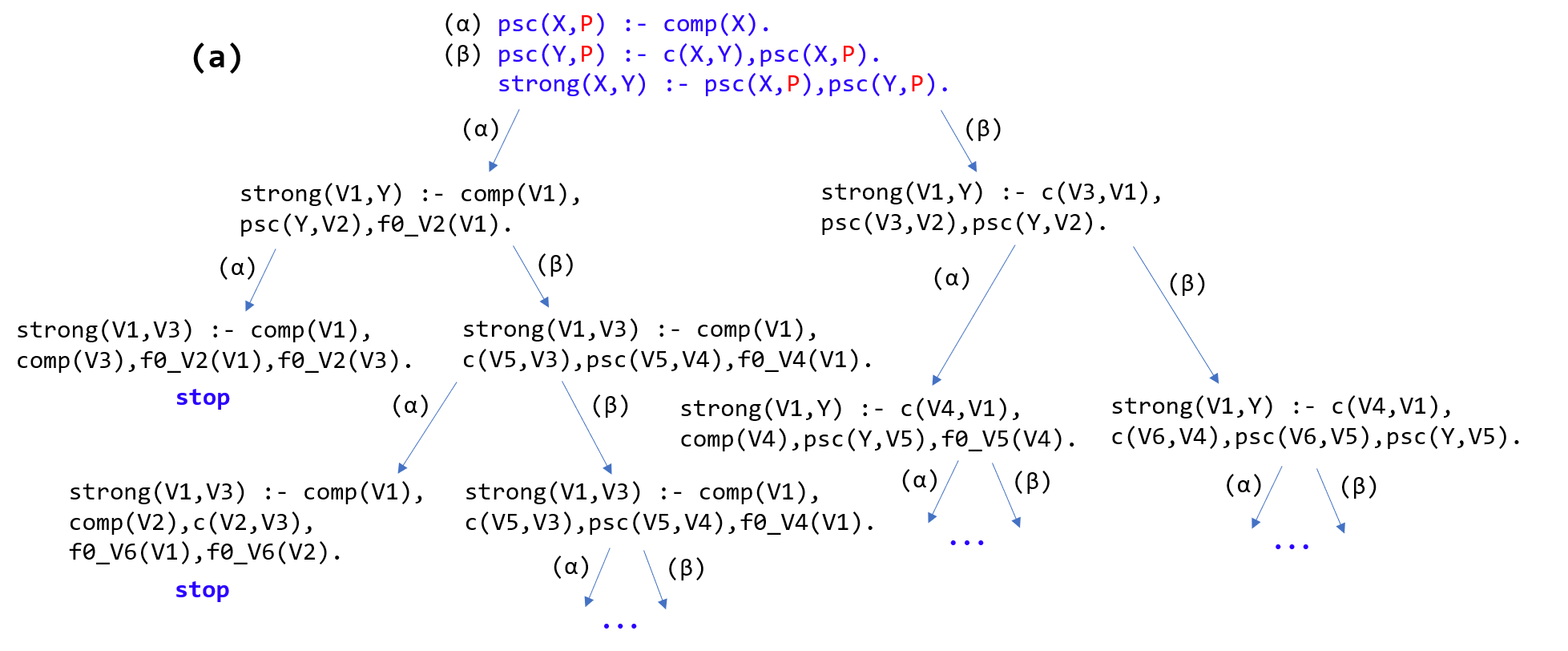}
    \caption{Role of Folding in HJE termination - without folding.}
    \label{fig:hjefoldingbeforebig}
    \vspace{3cm}
    \includegraphics[width=\textwidth]{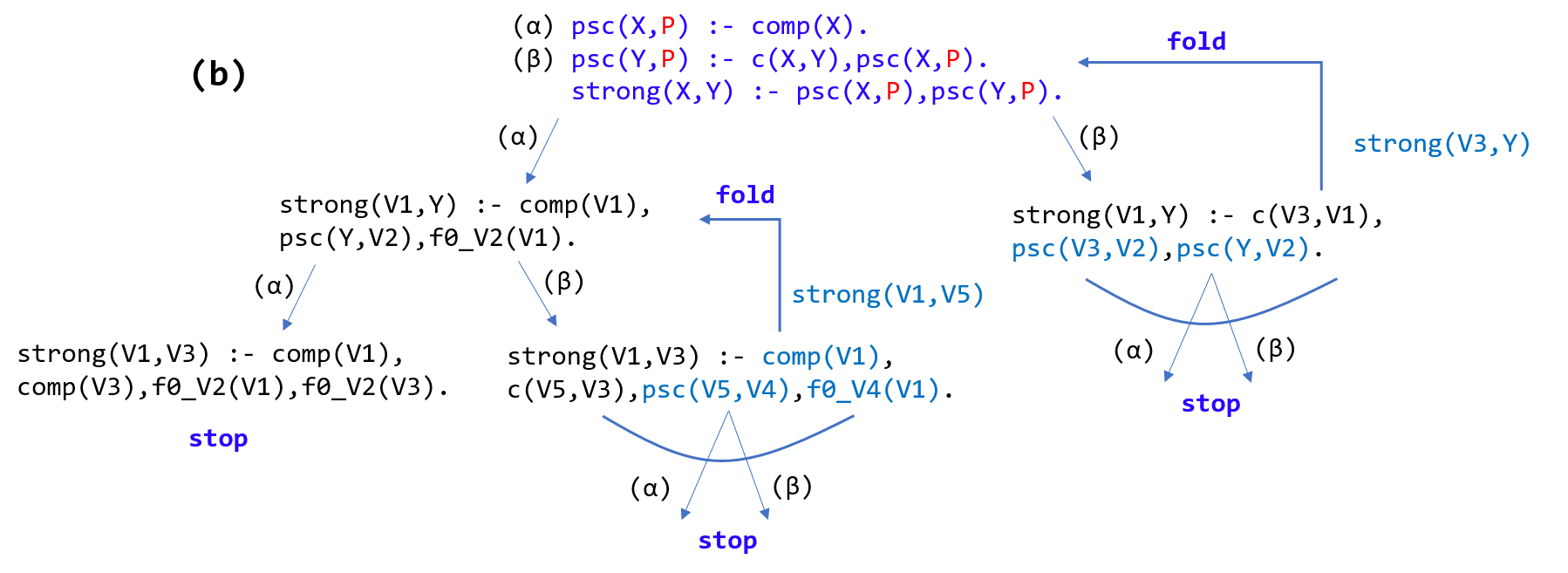}
    \caption{Role of Folding in HJE termination - with folding.}
    \label{fig:hjefoldingafterbig}
  \end{center}
\end{figure*}

\begin{figure*}[hbt!]
  \begin{center}
    \includegraphics[width=\textwidth]{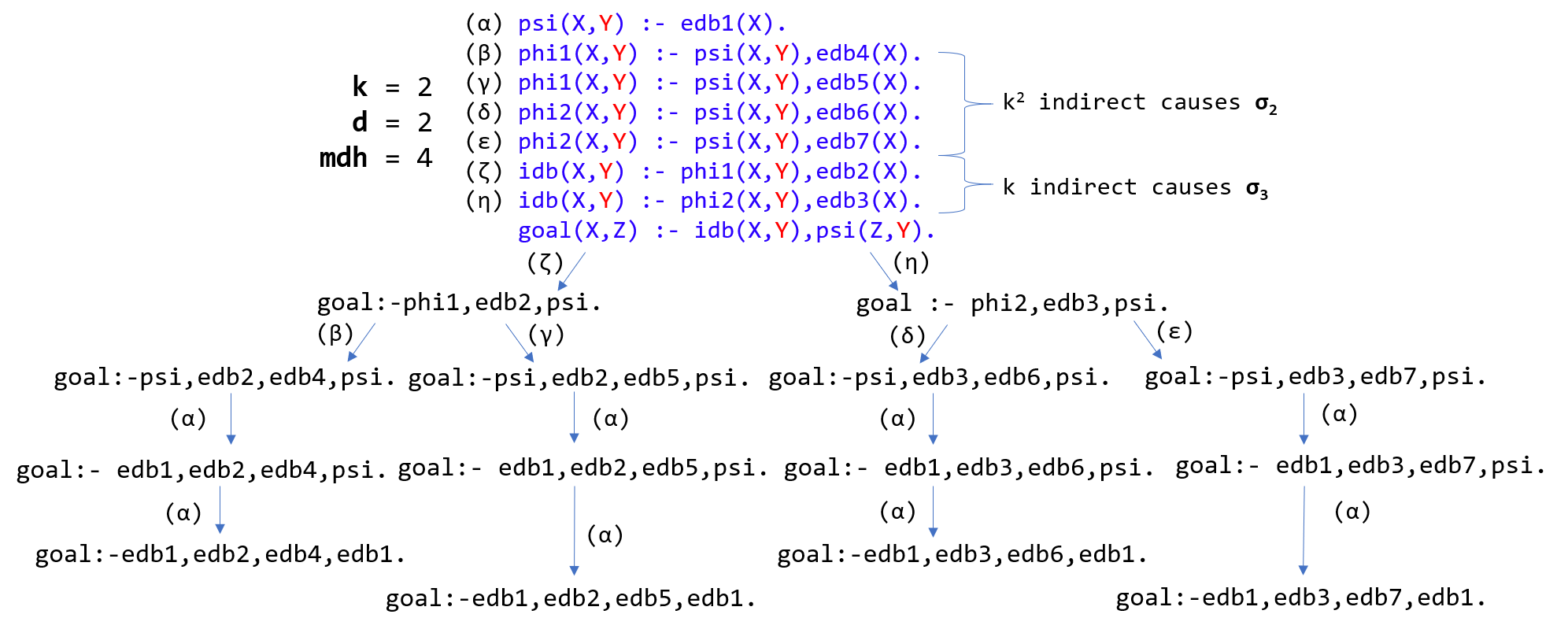}
    \caption{Relevant scenarios for HJE complexity - number of generated rules.}
    \label{fig:hjenumrulesbig}
    \vspace{3cm}
    \includegraphics[width=\textwidth]{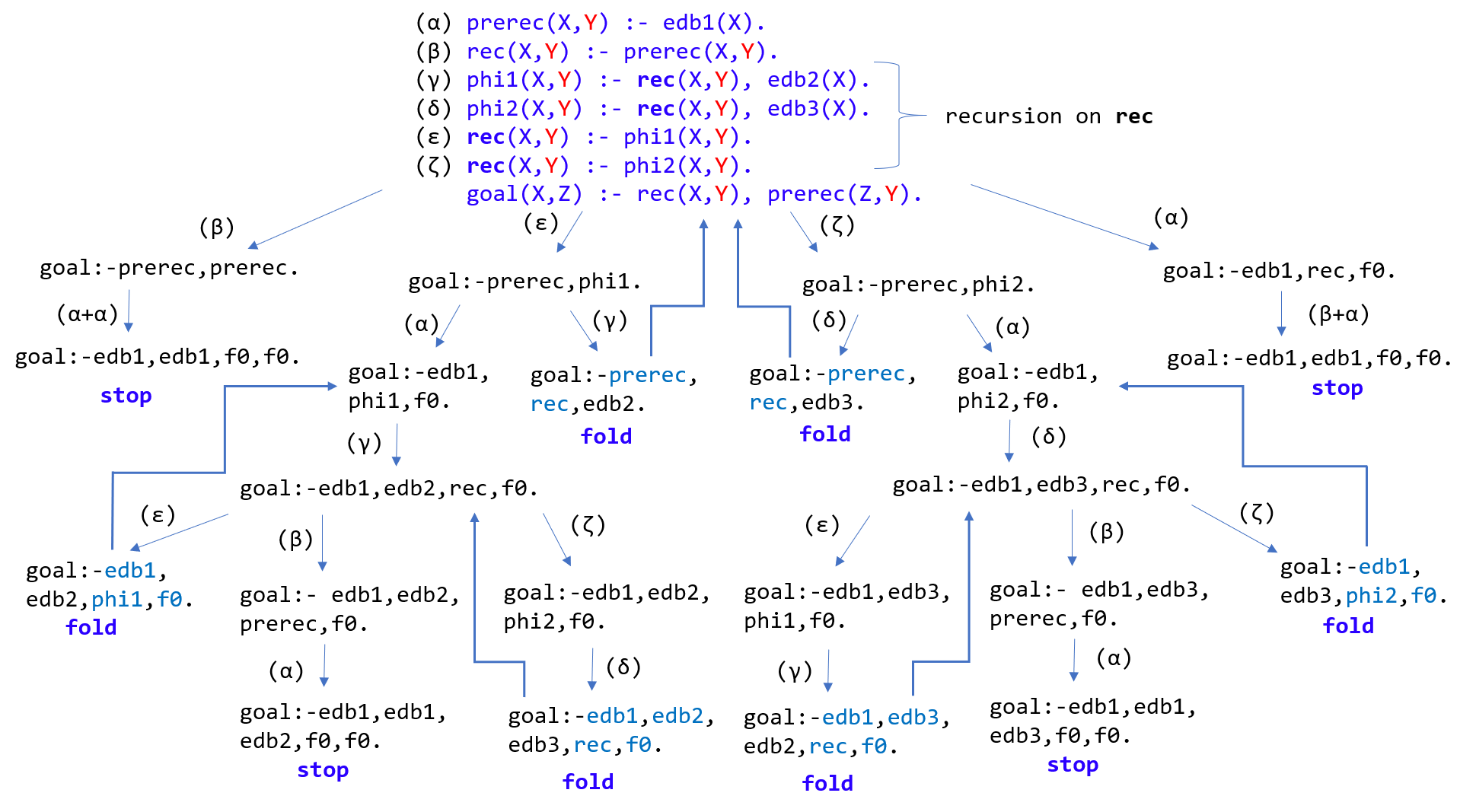}
    \caption{Relevant scenarios for HJE complexity - number of folding checks.}
    \label{fig:hjenumfoldsbig}
  \end{center}
\end{figure*}

\noindent
With reference to Section~\ref{hje} in the paper, we here provide a larger version of the figures illustrating relevant aspects of the Harmful Join Elimination algorithm (Figure~\ref{fig:hjefoldingbeforebig} to~\ref{fig:hjenumfoldsbig}). Moreover, Example~\ref{ex:hjefoldingresult} shows the normalized Harmless Warded set of rules after the application of the HJE algorithm to the set of rules in Figure~\ref{fig:hjefoldingbeforebig}.

\setcounter{example}{4}
\begin{example}
\label{ex:hjefoldingresult}
\begin{ttmath}
\begin{align*}
\textit{Original~rules~(causes~of~affectedness)}\\
1:psc(X,\overline P) :- \,\,comp(X).\\
2:psc(Y,\overline{P}) :- \,\,c(X,Y),psc(X,\overline{P}).\\
\textit{Rules~from~\texttt{Grounding}}\\
3:psc^\prime(X,\overline P) :- \,\,dom(\overline P),psc(X,\overline P).\\
4:psc(X,\overline P) :- \,\,psc^\prime(X,\overline P).\\
5:strong(X,Y) :- \,\,psc^\prime(X,\overline P),psc(Y,\overline P).\\
\textit{Rules~from~\texttt{Back-Composition}~with~\texttt{Skolem-Simplification}}\\
6:strong(V1,V1) :- \,\,comp(V1).\\
7:strong(V1,Y) :- \,\,c(V3,V1),strong(V3,Y).\\
8:strong(V1,V3) :- \,\,c(V5,V3),strong(V1,V5).
\end{align*}
\end{ttmath}
We briefly discuss the rules in the normalized set as output of HJE. Rules \texttt{1} and \texttt{2} are the remaining rules of the original set before normalization and, in this specific scenario, only correspond to the causes of affectedness for the harmful join rule removed. Rules \texttt{3},\texttt{4} and \texttt{5} are added during the Grounding phase of the procedure: as the atoms involved in the harmful join are isomorphic, these rules are only present once. Finally, the last three rules are the result of Back-Composition, via unfolding and folding (rule \texttt{6} and rules \texttt{7},\texttt{8}, respectively) and Skolem-Simplification, via linearization.
\end{example}

\newpage
\label{appendixHJE}
\section{Reasoning Evaluation}
\noindent

With reference to Section~\ref{testing} in the paper, we evaluate the performance of the Vadalog System on the structural scenarios previously presented, built with {\sc iWarded} and normalized with HJE.

\begin{figure}[!hb]
    \centering
    \includegraphics[width=0.45\textwidth]{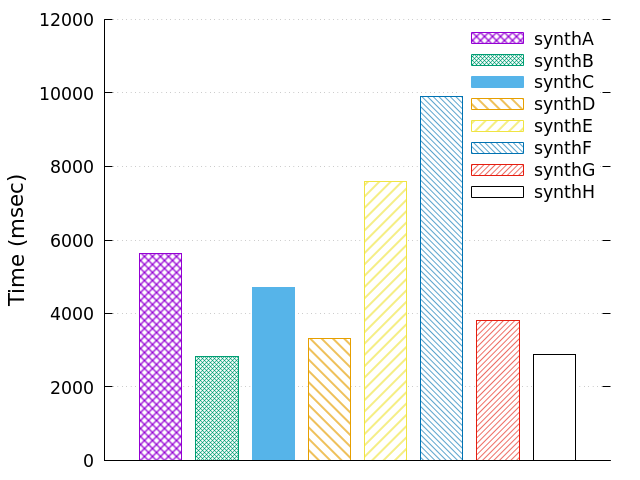}
    \caption{Reasoning Times for Structural Scenarios.}
    \label{fig:synthreasoning}
\end{figure}

\smallskip
\noindent
\textbf{Reasoning Times}. Figure~\ref{fig:synthreasoning} provides the reasoning times for the structural scenarios. The reasoner was here used as a ``library'' and invoked from specific Java test classes for end-to-end (storage to storage) reasoning. We used CSV files as storage, as they are supported by {\sc iWarded} and also allowed us to emphasize the performance of the reasoner: each input CSV file (built with {\sc iWarded} and one for each input atom) contains 1000 records for the corresponding input atom. It can be observed that the average performance of the reasoner is good: all the reasoning tasks are completed within 10 seconds. The results are comparable with the ones in~\cite{bellomarini2018vadalog}, to which we refer the reader for an in-depth explanation of the theoretical bases behind the impact the distinct features of the language, present in these scenarios (existentials, types of joins, recursions...), have on the reasoner. 

\begin{figure}[!hb]
    \centering
    \includegraphics[width=0.45\textwidth]{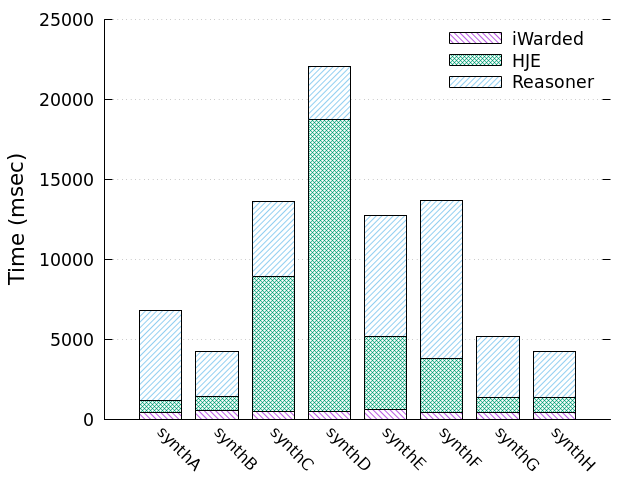}
    \caption{Complete Evaluation of Structural Scenarios.}
    \label{fig:synthcompleteeval}
\end{figure}

\smallskip
\noindent
\textbf{Complete Evaluation}. Figure~\ref{fig:synthcompleteeval} shows the generation times with {\sc iWarded}, the normalization times with HJE and the reasoning times over the structural scenarios, thus summarizing the distinct tests presented in this work involving the \textit{synths} and providing insight into a complete testing evaluation of realistic scenarios. As it can be observed, consistently with the sequence network abstraction and the memory-free generation approach presented in the paper, the generation of the sets of rules based on the distinct scenarios does not influence in a relevant manner the overall evaluation. On the other hand, the HJE applied to normalize \textit{synthC} and \textit{synthD} strongly affects the overall performance in the evaluation of these scenarios, which is expected due to the higher number of harmful join rules (20 and 50, respectively) and, consequently, the more probable presence of complex, possibly recursive, structures among the causes of affectedness: however, this cost is easily compensated, as the normalization of a set of rules only requires to be applied once, whereas the resulting harmless set is highly adaptable and reusable (as discussed in Section~\ref{iWarded}) for multiple reasoning tasks.

\newpage

\label{appendixtests}
\end{appendices}

\end{document}